\documentclass[10pt,
 prd,
 twocolumn,
 superscriptaddress,
 numerical,
 showpacs,
 amsmath,amssymb,
 aps,
nofootinbib,
longbibliography,
 floatfix
]{revtex4-1}

\usepackage{bm}
\usepackage{bbm}
\usepackage{bbold}
\usepackage{cancel}
\usepackage{slashed}
\usepackage{verbatim}
\usepackage{graphicx}
\usepackage{combelow} 
\usepackage{mathtools}
\usepackage[normalem]{ulem}
\usepackage[usenames,dvipsnames,svgnames,table]{xcolor}
\usepackage[colorlinks=True,linkcolor=blue,citecolor=blue,urlcolor=blue]{hyperref}

\definecolor{purple}{rgb}{0.8,0,0.6}
\definecolor{bluegreen}{rgb}{0.05,0.6,0.73}
\definecolor{forest}{RGB}{50,150,50}
\newcommand{\corr}[1]{#1}

\allowdisplaybreaks
\begin{document}

\title{Enhanced Condensation Through Rotation}

\author{Maxim Chernodub}\thanks{Corresponding author (email: \href{mailto:maxim.chernodub@univ-tours.fr}{maxim.chernodub@univ-tours.fr}).}
\affiliation{Institut Denis Poisson, CNRS UMR 7013, Universit\'e de Tours, Universit\'e d'Orl\'eans, 
Parc de Grandmont, Tours, 37200, France}
\affiliation{Department of Physics, West University of Timi\cb{s}oara, Bd.~Vasile P\^arvan 4, Timi\cb{s}oara 300223, Romania}

\author{Frank Wilczek}
\affiliation{Center for Theoretical Physics, Massachusetts Institute of Technology, Cambridge, Massachusetts 02139, USA}
\affiliation{T. D. Lee Institute, Shanghai 201210, China}
\affiliation{Wilczek Quantum Center, Department of Physics and Astronomy, Shanghai Jiao Tong University, Shanghai 200240, China}
\affiliation{Department of Physics and Origins Project, Arizona State University, Tempe, Arizona 25287, USA}
\affiliation{Department of Physics, Stockholm University, AlbaNova University Center, 106 91 Stockholm, Sweden}
\affiliation{Nordita, Stockholm University and KTH Royal Institute of Technology, Hannes Alfv\'{e}ns v\"{a}g 12, SE-106 91 Stockholm, Sweden}

\begin{abstract}
We argue that rotation of a thin superconducting cylinder can increase the critical superconducting temperature substantially. A purely rotational effect originates from the tendency of a steadily rotating mechanical system to maximize its moment of inertia. A condensation of Cooper pairs in a rotating cylinder decouples a part of the normal electron fraction from rotation, thus producing a circulating electric current of an uncompensated electric charge of lattice ions. The current generates the magnetic field that stores energy of rotation, thus increasing the moment of inertia. In the presence of an external magnetic field, another enhancement effect originates from the interaction energy of the dipole magnetic moment of the normal component with the background magnetic field. In both cases, rotation of the cylindrical shell promotes the formation of condensate that decouples from mechanical rotation. We give quantitative estimates for a thin cylinder of aluminum.  
\end{abstract}

\date{\today}

\maketitle

\section{Introduction}

The core observation we make here is very simple, when stated naively. Motivated by a ``two-fluid'' picture, one might expect that a superconducting condensate decouples from the rotational motion of the normal component. The residual density of moving charge density results in a current and thus a magnetic field.  In the presence of a properly oriented background magnetic field, this can lead to cancellation that decreases the total magnetic energy.
Thus, it becomes energetically advantageous to put more substance into the condensate, which increases the critical temperature. Even without the external field, the magnetic field contributes to the moment of inertia of the system, which lowers the free energy in the corotating reference frame; this effect also encourages condensate formation.  As we shall discuss, these simple motivating thoughts, after appropriate revision and qualification, retain important elements of truth.

Below we review and extend the theory of the rotating superconductors, and show that the rotation of a thin superconducting cylindrical shell can catalyze the emergence of superconductivity. The effect is facilitated by the mismatch of the normal and supercurrent velocities.

Consider a solid superconducting cylinder rotating uniformly with a constant angular velocity $\boldsymbol{\Omega}$ about its symmetry axis. At zero temperature, all electrons form Cooper pairs and condense into a charged superfluid, which interacts with a rotating, positively charged ionic lattice.  In the absence of mechanical friction between the ionic lattice and the charged superfluid condensate, one might naively argue that the superfluid component would remain in a static, non-rotating state to minimize its kinetic energy. Such behavior would be analogous to the lack of rotational response expected of a neutral superfluid confined within a very slowly rotating vessel.  Here, however, 
the rotation of the crystal lattice induces a circular electric current of positively charged ions. This current produces a magnetic field along the rotation axis, perceived by the charged superfluid as an external background field.  The magnetic field generated by the rotating crystal arises intrinsically from within the bulk at every point of the superconductor.

To mitigate the effect of this energetically costly bulk magnetic field, which frustrates superconductivity, the condensate produces a Meissner supercurrent. In this way, the negatively charged superfluid fraction synchronizes its velocity with the velocity of the positively charged crystal lattice, ensuring that in the bulk of the superconductor, the total electric current vanishes. Thus, even in the absence of a pho{\underline n}on-mediated coupling between the rotating ionic lattice and the condensate, the rotation rigidly drags the charged superfluid via a pho{\underline t}on-mediated interaction in bulk.

Still, a rotating superconductor, regardless of its chemical composition, develops the bulk magnetic field (which is also called the ``London magnetic field'')~\cite{London1961, Becker1933}:
\begin{align}
    {\boldsymbol{B}}_L = \frac{2 m c}{e} {\boldsymbol{\Omega}}\,.
    \label{eq_B_L}
\end{align}
This field is generated by a surface layer of the cylinder, where the velocities of the normal and condensed electronic fractions differ from each other~\cite{Capellmann2002}.  With this, the bulk vector potential relieves the potential for frustration associated with non-zero vorticity of the superflow.

The appearance of \corr{the London magnetic field, given by Eq.~\eqref{eq_B_L}}, in the bulk of a uniformly rotating London superconductor is a natural consequence of the rigid drag of the Cooper pairs by the rotating ionic lattice: the system tends to minimize the energy of the magnetic field that would appear in the bulk of the superconductor if Cooper pairs were decoupled from rotation. For a thin, hollow cylindrical shell, this dragging mechanism does not work. Therefore, the charged superfluid component decouples from the rotational mechanical motion of the cylinder. In our article, we show that this decoupling strongly enhances superconductivity and raises the critical temperature of the superconducting transition.

The paper is organized as follows. In Section~\ref{sec_thin_cylinder} we derive the free-energy functional for the superconducting order parameter of a thin superconducting cylindrical shell that rotates uniformly about its axis. In Sec.~\ref{sec_phase}, we analyze how the superconducting critical temperature is altered by rotation in the absence of an external magnetic field and then examine the combined effects arising from the interplay between rotation and a background field coaligned with the rotation axis. Section~\ref{sec_strength} is devoted to the estimation of the strength of the effects in a realistic system. This section also discusses other potentially relevant phenomena and shows that they are subleading. The last section summarizes our conclusions. Appendix~\ref{sec_comments} confronts our analysis with earlier studies.

\section{Rotating thin cylinder}
\label{sec_thin_cylinder} 

\subsection{The system}

\begin{figure}[!htb]
    \centering
    \includegraphics[width=0.5\linewidth]{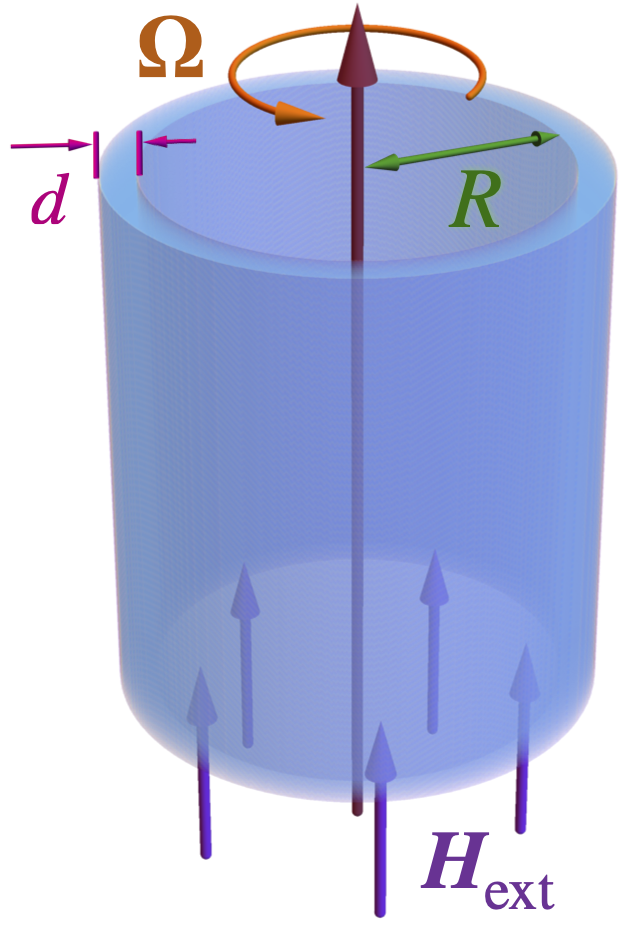}
    \caption{Setup: a thin cylinder with the radius~$R$ made of a superconducting film of the thickness $d \ll R$ rotating with the constant angular velocity $\boldsymbol{\Omega}$ in the background of the external magnetic field ${\boldsymbol{H}}_{\rm ext} \| {\boldsymbol{\Omega}}$.}
    \label{fig_cylinder}
\end{figure}

Consider a hollow cylinder made of a thin superconducting film as shown in Fig.~\ref{fig_cylinder}. Following the Little-Parks setup~\cite{Little1962}, we consider a thin superconducting film of a thickness $d$ deposited on a cylindrical insulator of a radius $R \gg d$ and a height $L_z$ in an external magnetic field. If the thickness of the film $d$ is smaller than the London penetration length, $\lambda_L$, then the rotation of the ionic lattice produces a negligible Meissner current, and the kinetic energy of the condensate can be neglected. In this case, at a finite temperature $T$ below the superconducting phase transition, $T < T_c$, the electrons are shared between the condensate and the normal electron component. In addition to requiring $d \lesssim \lambda_L$, we take the thickness of the film to be smaller than the coherence length, $d \lesssim \xi$ implying that the absolute value of the order parameter $|\psi|$ is a spatially homogeneous quantity.  Spatial dependence of the condensate appears only in its phase: $\psi({\boldsymbol{x}}) = |\psi| e^{i \theta({\boldsymbol{x}})}$~\cite{DeGennes2018}. 

The Ginzburg-Landau approach to rotating superconductors has an extensive literature~\cite{Verkin1972, Capellmann2002, Berger2004, Lipavsk2013, Babaev2014, Hirsch2019a, Hirsch2019b}. In our article, we reexamine the energy balance for a thin superconducting cylinder rotating in the background of a magnetic field $\boldsymbol{H}_{\rm ext}$. We emphasize the importance of two key facts that favor condensation through rotation: first, that normal and superconducting electrons share a common reservoir of charge carriers; and second, that the normal component contributes significantly to the interaction energy of the magnetic dipole moment of the rotating superconductor in the background magnetic field.

The total free energy of a rotating superconductor,
\begin{align}
    F = F_{\rm supr} + F_{\rm mech} + F_{\rm magn}\,,
    \label{eq_F_total_all}
\end{align}
is a sum of the contributions coming from the superconducting condensate, $F_{\rm supr}$, the classical mechanical motion of the non-superconducting electronic component $F_{\rm mech}$ in the co-rotating reference frame, and the magnetic field generated by the circular motion of the electrically charged normal constituent, $F_{\rm magn}$, respectively. Below, we analyze these contributions separately.

\subsection{Free energy of the superconducting condensate}

The Ginzburg-Landau (GL) free energy of the superconducting condensate $\psi = \psi({\boldsymbol{x}})$ is~\cite{DeGennes2018}:
\begin{align}
    F_{\rm supr} = \int_{V_{\rm s}} d^3 x \biggl[\frac{1}{4 m} 
    {\Bigl| \Bigl(\frac{\hbar}{i} {\boldsymbol{\nabla}} 
    + \, \frac{2 e}{c} {\boldsymbol{A}}  \Bigr) \psi \Bigr|}^2 \nonumber\\
    +  \alpha |\psi|^2 + \frac{\beta}{2} |\psi|^4 \biggr]\,,
\label{eq_F_GL}
\end{align} 
where $\boldsymbol{A}$ is the gauge field, and $\alpha$ and $\beta > 0$ are the GL parameters.\footnote{\label{foot_phonons} 
For a rotating superconductor, the GL energy~\eqref{eq_F_GL} can also be extended by an additional non-dissipative term that accounts for the effect of rotation on the phonon-mediated coupling responsible for the formation of the Cooper pairs. We omit this term because it leads only to a modification of the superconducting current via a renormalized electron mass ($m \to m^*$)~\cite{Capellmann2002}. As we show below, the Meissner current, generated by the rotation of the cylinder, plays a negligible role in our geometry.
} 
The superconductivity is supported by a finite density of Cooper pairs:
\begin{align}
n_{\rm s} = |\psi|^2\,. 
\label{eq_n_s}
\end{align}
Each pair has a mass $2 m$ and an electric charge twice that of an electron, $-2e$, with $m = m_e$ and $e = |e| > 0$. The integral in Eq.~\eqref{eq_F_GL} is taken over the whole volume 
\begin{align}
	V_{\rm s} = 2 \pi R L_z d\,, 
\end{align}
of the thin cylindrical superconducting film.  

\corr{We employ the minimal GL functional~\eqref{eq_F_GL}, whose quartic double-well potential provides the standard description of a system undergoing a continuous transition between the normal and superconducting phases. For a more quantitative description of realistic materials, the potential may be extended by including higher-order terms, such as $|\psi|^6$, $|\psi|^8$, and higher powers.}

Below, we use cylindrical coordinates ${\boldsymbol{x}} = (r , \varphi, z)$ with the symmetry axis of the cylinder pointing along the $z$ direction. We ignore geometrical $O\bigl(d/R\bigr)$ subleading corrections that arise due to a finite film thickness $d \ll R$.

\subsection{Mechanical free energy}

The mechanical energy given by the second term in Eq.~\eqref{eq_F_total_all}, corresponds to the sum of the rotational kinetic energies of the ions in the crystal lattice  ($\ell = {\rm I}$), electrons in the normal state ($\ell = {\rm n}$), and the Lagrange term that couples the total angular momentum density of the system ${\boldsymbol{l}} = {\boldsymbol{l}}_{\rm I}({\boldsymbol{x}}) + {\boldsymbol{l}}_{\rm n}({\boldsymbol{x}})$ with the angular velocity $\boldsymbol{\Omega}$:
\begin{align}
    F_{\rm mech} =  \int_{V_{\rm s}} d^3 x \sum_{\ell = {\rm I, n}}\Bigl(\frac{1}{2} \rho_\ell {\boldsymbol{v}}^2_\ell 
    \,-\, \boldsymbol{\Omega\cdot {\boldsymbol{l}}_{\ell}}\Bigr)\,.
\label{eq_F_I_general}
\end{align}
Here $\rho_{\rm n}$ and $\rho_{\rm I}$ are the mass densities of the normal fraction of electrons and the ionic lattice. The total angular momentum density gets contributions from the ionic lattice and the normal electron fraction, ${\boldsymbol{l}}_\ell = {\boldsymbol{x}} \times {\boldsymbol{p}}_\ell$, where ${\boldsymbol{p}}_\ell = m_\ell {\boldsymbol{v}}_\ell$ are the corresponding momenta with $\ell = {\rm I}, {\rm n}$. The local velocity $\boldsymbol{v}_{\rm n}$ of the normal fraction of electrons and the velocity of the ionic lattice $\boldsymbol{v}_{\rm I}$ coincide 
\begin{align}
    \boldsymbol{v}_{\rm n} ({\boldsymbol{x}}) = \boldsymbol{v}_{\rm I} ({\boldsymbol{x}}) = \boldsymbol{v}({\boldsymbol{x}}) = {\boldsymbol{\Omega}} \times {\boldsymbol{x}}\,,
    \label{eq_v_n}
\end{align}  
because the phonon-mediated interaction synchronizes their rotational motion in thermal equilibrium. 

The mechanical rotational energy of the normal-state electrons can be inferred by noticing that both normal and superconducting electrons share a common reservoir. Consequently, the number density of normal electrons $n_{\rm n}$ is directly related to the number density of the superconducting Cooper pairs $|\psi|^2$: a stronger superconducting condensate leads to a reduced population of normal electrons and vice versa. To calculate the density of normal electrons, we notice that in thermal equilibrium, the superconductor is electrically neutral at every point. Therefore, the electric charge density of the superconducting component, $- 2 e |\psi|^2$, is compensated by the sum of the charge densities of the normal component, $-e n_{\rm n}$, and the ionic lattice, $+ e Z_{\rm I} n_{\rm I}$. The latter is expressed via the number density of ions, $n_{\rm I}$, and the electric charge of each ion, $+e Z_{\rm I}$. Then, the local neutrality condition, $- 2 e |\psi(\boldsymbol{x})|^2 - e n_{\rm n} + e Z_{\rm I} n_{\rm I} = 0$, gives us the number density of the normal electrons, $n_{\rm n} (\boldsymbol{x}) = Z_{\rm I} n_{\rm I} - 2 |\psi(\boldsymbol{x})|^2$, as well as their mass density:
\begin{align}
	\rho_{\rm n} (\boldsymbol{x}) \equiv m n_{\rm n} (\boldsymbol{x}) = m\bigl( Z_{\rm I} n_{\rm I} - 2 |\psi(\boldsymbol{x})|^2\bigr)\,.
    \label{eq_rho_n}
\end{align}
The local number density of ions, $n_{\rm I}$, does not depend on the angular velocity $\boldsymbol{\Omega}$ since the non-relativistic rotation does not deform the ionic lattice. Thus, the mass density of the ions is a constant quantity, $\rho_{\rm I}({\boldsymbol{x}}) = m_{\rm I} n_{\rm I}$, where $m_{\rm I}$ is an effective mass of an ion in the crystal. 

Equations~\eqref{eq_F_I_general}, \eqref{eq_v_n} and \eqref{eq_rho_n} provide us with the rotational energy of the normal component in the corotating reference frame:
\begin{align}
    F_{\rm mech} = - \frac{1}{2} I_{\rm mech} \Omega^2\,,
\label{eq_F_I}
\end{align}
where
\begin{align}
    I_{\rm mech} = I_{\rm n0} + I_{\rm s}\,,
	\label{eq_I_n} 
\end{align}  
is the moment of the mechanically rotation part of the system given by the ions and the normal electron fraction. Here,
\begin{align}
    I_{\rm n0} =  (m_{\rm I} + m Z_{\rm I}) n_{\rm I}  R^2 V_{\rm s}\,,
	\label{eq_I_n0}
\end{align}
corresponds to the moment of inertia of the film cylinder in the absence of the condensate, $\psi = 0$, if all electrons were in the normal state.

The second term in Eq.~\eqref{eq_I_n} is the contribution coming from the electrons in the condensed Copper pairs. Remarkable properties of this term are that the emergent effective classical moment of inertia~$I_{\rm s}$ (i) depends explicitly on the quantum superconducting condensate $\psi$ and (ii) has a negative value in the superconducting state with $\psi \neq 0$: 
\begin{align}
	I_{\rm s} = - 2 m \int_{V_{\rm s}} d^3 x \, r^2 |\psi(\boldsymbol{x})|^2 \,.
    \label{eq_I_s}
\end{align}

A negative contribution to the moment of inertia coming from the condensate of Cooper pairs~\eqref{eq_I_s}, $I_{\rm s} < 0$, has the transparent physical sense: the larger the density of the superconducting pairs, the lower the total rotational energy carried by electrons in the normal fraction\footnote{The negative contribution to the superconducting fraction of electrons to the total moment of inertia, Eqs.~\eqref{eq_I_n} and \eqref{eq_I_s}, does not in any way imply that the electrons in the superconducting state possess a negative mass. On the contrary, the mass of a Cooper pair is a positive quantity. A negative moment of inertia has also been found in the numerical simulations of a completely different physical system, a hot gluon plasma~\cite{Braguta:2023yjn}.}.

\subsection{Free energy associated with the magnetic field}
\label{sec_free_magnetic_energy}

\subsubsection{Magnetic field generated by rotation}

In the presence of the superconducting condensate, the normal component ---that comprises both the normal electrons and the ionic lattice--- has a nonvanishing charge density. Using Eq.~\eqref{eq_v_n} and the condition of the local charge neutrality, one can show that the circular motion of electric charges, associated with the rotation of the ionic lattice and the normal fraction of electrons, ${\boldsymbol{J}}_{\rm n} = e Z_{\rm I} n_{\rm I} \boldsymbol{v}_{\rm I} - e n_{{\rm n}} \boldsymbol{v}_{\rm n}$, generates a circular \corr{normal} electric current proportional to the superconducting density: 
\begin{align}
    {\boldsymbol{J}}_{\rm n}({\boldsymbol{x}}) 
    = 2 e |\psi({\boldsymbol{x}})|^2 ({\boldsymbol{\Omega}} \times {\boldsymbol{x}})\,.
    \label{eq_J_normal}
\end{align}

\corr{There are three fascinating facts associated with the current~\eqref{eq_J_normal}. First, for the superconducting electrons, the normal current~\eqref{eq_J_normal} appears as an external current generated from within the whole volume of the rotating superconductor. This current is carried by the mechanically rotated ionic lattice that drags with it the normal electron fraction.}

\corr{Second, the normal current~\eqref{eq_J_normal}, despite it appears inside the superconductor, is controlled externally by the mechanical rotation of the superconducting crystal.}

\corr{Third, the normal current~\eqref{eq_J_normal} is proportional to the density of the superconducting condensate $|\psi|^2$. As we mentioned earlier, this surprising feature appears as a result of the charge neutrality: the superconducting fraction (that has the electric charge density $-2 e |\psi|^2$) decouples from rotation, and the only rotating part of the system is given by the normal fraction that has, dictated by the charge neutrality, the electric charge density $+2 e |\psi|^2$. The neutrality property makes the normal current~\eqref{eq_J_normal} naturally proportional to $|\psi|^2$.}

The current density of the normal component~\eqref{eq_J_normal} should be distinguished from the current of Cooper pairs:
\begin{align}
	    {\boldsymbol{J}}_{\rm s} = &\, - \frac{1}{c} \frac{\delta F_{\rm supr}}{\delta {\boldsymbol{A}}} 
        \nonumber \\
        = &\, \frac{i \hbar e}{2m} \bigl(\psi^* {\boldsymbol{\nabla}} \psi - \psi {\boldsymbol{\nabla}} \psi^* \bigr) - \frac{2e^2}{mc} {\boldsymbol{A}} |\psi|^2 \,.
    \label{eq_J_superconducting}
\end{align}

The last term in the free energy~\eqref{eq_F_total_all} is given by the energy associated with the magnetic degrees of freedom:
\begin{subequations}
\begin{align}
    F_{\rm magn} = &\, \frac{1}{8 \pi} \int d^3 x\, {\boldsymbol{B}}^2 
\label{eq_F_magn_B2}\\
    &\, - \frac{1}{4 \pi} \int d^3 x\, {\boldsymbol{B}} \cdot \boldsymbol{H}_{\rm ext} 
\label{eq_F_magn_BH}\\
    & \, - \frac{1}{c} \int d^3 x\, {\boldsymbol{A}} \cdot \boldsymbol{J}_{\rm n}\,.
\label{eq_F_magn_AJ}
\end{align}
    \label{eq_F_magn}
\end{subequations}
The first term~\eqref{eq_F_magn_B2} is the energy stored in the generated magnetic field, $\boldsymbol{B} = \boldsymbol{\nabla} \times \boldsymbol{A}$. The second term~\eqref{eq_F_magn_BH}
is the Lagrange multiplier that encodes the interaction of the magnetic field $\boldsymbol{B}$ with the externally controlled background magnetic field $\boldsymbol{H}_{\rm ext}$. The external magnetic field is produced by the current ${\boldsymbol J}_{\rm ext}$, which circulates far from the superconductor, 
\begin{align}
	{\boldsymbol{\nabla}} \times {\boldsymbol{H}}_{\rm ext} = \frac{4\pi}{c} {\boldsymbol J}_{\rm ext}\,.
\label{eq_Hext_J}
\end{align}
Finally, the third term~\eqref{eq_F_magn_AJ} is the standard electromagnetic interaction of the normal electric current ${\boldsymbol{J}}_{\rm n}$ produced by the rotating cylinder with the electric field represented by the gauge potential~$\boldsymbol{A}$. The first two integrals in Eq.~\eqref{eq_F_magn} are taken over the entire space because the magnetic field extends beyond the superconductor, while the third integral is taken over the interior of the film that supports the normal current $\boldsymbol{J}_n$.

The field $\boldsymbol{B}$, produced by all electric currents in the system, can be found via the Amp\`ere law, which corresponds to one of the Ginzburg-Landau equations:
\begin{align}
	{\boldsymbol{\nabla}} \times {\boldsymbol{B}}({\boldsymbol{x}}) = \frac{4 \pi}{c} ({\boldsymbol J}_{\rm n} + {\boldsymbol J}_{\rm s} + {\boldsymbol J}_{\rm ext})\,. 
    \label{eq_nabla_B}
\end{align}
This equation is a result of a variation of the full free energy~\eqref{eq_F_total_all}, including the magnetic energy~\eqref{eq_F_magn}, with respect to the gauge potential~$\boldsymbol{A}$. The right-hand side of Eq.~\eqref{eq_nabla_B} shows that the source of $\boldsymbol{B}$ is given by all electric currents in the system, given by the sum of the normal~\eqref{eq_J_normal}, superconducting~\eqref{eq_J_superconducting} and external~\eqref{eq_Hext_J} electric currents, respectively.

For a cylindrical shell made of a very thin film of a thickness $d$ and a macroscopic radius $R \gg d$ with a uniform amplitude of the condensate $|\psi|$, the current density of the normal component~\eqref{eq_J_normal} is\footnote{A uniform condensate inside a cylindrical film of a thickness $d$ and a radius $R$ is $|\psi({\boldsymbol{x}})| \equiv |\psi(r)| = |\psi| \Theta(R + d - r) \Theta(r-R)$, where $\Theta = \Theta(x)$ is a Heaviside step function. In the limit $R \gg d$, one can use the approximation $|\psi({\boldsymbol{x}})| \to d \, \delta(r - R)$.} 
\begin{align}
    {\boldsymbol{J}}_{\rm n}({\boldsymbol{x}}) = 2 e |\psi|^2 \Omega R d\, \delta(r  - R) {\bf e}_\varphi\,, 	
    \label{eq_J_normal_cyl} 
\end{align}
where ${\bf e}_\varphi$ is a polar vector. The superconducting current density~\eqref{eq_J_superconducting}, 
\begin{align}
    {\boldsymbol{J}}_{\rm s} = - 2 e |\psi({\boldsymbol{x}})|^2 {\boldsymbol{v}}_{\rm s} \,,	
    \label{eq_J_superconducting_cyl_0}
\end{align}
is proportional to the superfluid velocity, $ {\boldsymbol{v}}_{\rm s} = v_s {\bf e}_\varphi$:
\begin{align}
    {\boldsymbol{v}}_{\rm s} = \frac{1}{2m} \Bigl(\hbar {\boldsymbol{\nabla}} \theta + \frac{2e}{c} {\boldsymbol{A}}\Bigr) = \frac{\hbar}{2 m R} \Bigl(k + {\bar \phi}_B \Bigr) {\bf e}_\varphi 
     \,,
    \label{eq_v_s}
\end{align}
which depends on the winding number $k \in {\mathbb Z}$ of the phase of the condensate, $\theta \equiv {\rm arg}\,\psi = k \varphi$, and the magnetic flux $\phi_B = \pi R^2 B$ of the total (external and generated by rotation) magnetic field threading the cylinder. The total magnetic flux is expressed in the dimensionless form as ${\bar \phi}_B = \phi_B / \phi_0$, where~\cite{Kittel2018} 
\begin{align}
	\phi_0 = \frac{2\pi\hbar c}{2e} \simeq 2.068\times 10^{-7}\, {\rm G} \, {\rm cm}^2\,,
    \label{eq_phi_0}
\end{align}
denotes the magnetic flux quantum. Using Eq.~\eqref{eq_v_s}, one can represent the superconducting current~\eqref{eq_J_superconducting_cyl_0} as 
\begin{align}
    {\boldsymbol{J}}_{\rm s} = - \frac{e \hbar}{m R} |\psi|^2  \Bigl(k + {\bar \phi}_B \Bigr) d \, \delta(r  - R) {\bf e}_\varphi\,.
    \label{eq_J_superconducting_cyl}
\end{align}

As we argue below, the normal-current component~\eqref{eq_J_normal_cyl} dominates over the superconducting current~\eqref{eq_J_superconducting_cyl} in our geometry, $|\boldsymbol{J}_{\rm s}| \ll |\boldsymbol{J}_{\rm n}|$. Therefore, we evaluate the magnetic free energy~\eqref{eq_F_magn} taking into account only the contribution of the circulating electric current corresponding to the normal component~\eqref{eq_J_normal_cyl}.

The circular normal electric current~\eqref{eq_J_normal} of ions and unpaired normal electrons produces the uniform magnetic field $\boldsymbol{B}_{\rm n} = B_{\rm n} {\bf e}_z$ with the strength
\begin{align}
B_{\rm n}(r) = &\, \frac{4 \pi \sigma}{c} R \Omega \, \Theta(R - r) \nonumber\\
\equiv &\, \frac{8 \pi e}{c} |\psi|^2 R d \, \Omega \, \Theta(R - r) \,,
\label{eq_B_Omega}
\end{align}
where $\sigma = \rho d \equiv 2 e d |\psi|^2$ is the surface density of the electric charge carried by the normal component that is by the ions and the neutral electron fraction. The magnetic field~\eqref{eq_B_Omega} appears only in the interior of the cylinder, $r \leqslant R$, supported by the Heaviside $\Theta$ function. Ignoring the superconducting current ${\boldsymbol{J}}_{\rm s}$, we get the solution of the Amp\`ere law:
\begin{align}
	{\boldsymbol{B}}(r) = B(r)\, {\bf e}_z\,, 
    \qquad 
    B(r) = B_{\rm n}(r) + H_{\rm ext}\,,
    \label{eq_B_r}
\end{align}
with the gauge field:
\begin{align}
	{\boldsymbol{A}}(r) = \frac{1}{2} r B(r) \, {\bf e}_\varphi\,.
    \label{eq_A_r}
\end{align}

Taking Eqs.~\eqref{eq_J_normal_cyl}, \eqref{eq_B_Omega}, \eqref{eq_B_r} and \eqref{eq_A_r} to the magnetic free energy~\eqref{eq_F_magn} and ignoring the constant energy density of the background magnetic field, ${\boldsymbol{H}}^2_{\rm ext}/(8 \pi)$, we get 
\begin{align}
	F_{\rm magn} = F_{B} + E_{\rm dipole}\,.
    \label{eq_magn_B_E}
\end{align}

\subsubsection{Free energy of generated magnetic field}

The first term in Eq.~\eqref{eq_magn_B_E} is the free energy of the generated magnetic field ${\boldsymbol{B}}_n$:
\begin{align}
	F_{B} = - \frac{1}{8 \pi} \int d^3 x\, {\boldsymbol{B}}^2_n = - \frac{1}{2} I_{\rm magn} \Omega^2\,,
    \label{eq_F_B}
\end{align} 
where the quantity
\begin{align}
	I_{\rm magn} = \frac{8 \pi e^2}{c^2} d R^3 \int d^3 x\, |\psi(\boldsymbol{x})|^4 \,,
    \label{eq_I_magn}
\end{align}
is interpreted as the moment of inertia associated with the generated magnetic field. 

\corr{The free energy~\eqref{eq_F_B} is given by a sum of the first term~\eqref{eq_F_magn_B2} and the third term~\eqref{eq_F_magn_AJ} in the magnetic free energy~\eqref{eq_F_magn}. Its particularly simple form is justified in our parametric regime when the superconducting current ${\boldsymbol{J}}_{\rm s}$ is neglected as compared to a much stronger normal current ${\boldsymbol{J}}_{\rm n}$ [{\it cf.} the discussion around Eq.~\eqref{eq_Jn_dominance} below].}

\corr{The normal current generates the uniform magnetic field $({\boldsymbol{\nabla}} \times {\boldsymbol A}) = \boldsymbol{B}_{\rm n} = B_{\rm n} {\bf e}_z$ of the constant strength $B_{\rm n}$ given in Eq.~\eqref{eq_B_Omega}. Setting, for convenience, the background field to zero, ${\boldsymbol{H}}_{\rm ext} = 0$, and choosing the gauge potential as ${\boldsymbol{A}} = A_\varphi(r) {\bf e}_\varphi$, one finds a solution for the gauge potential: $A_\varphi(r) = B_{\rm n} r/2$ inside the cylinder ($r \leqslant R$) and $A_\varphi(r) = B_{\rm n} R^2/(2 r)$ outside it ($r > R$).}

\corr{The third term in the magnetic free energy~\eqref{eq_F_magn} reduces to the integral over the volume of the thin superconducting shell because the normal current ${\boldsymbol{J}}_{\rm n}$ takes a non-zero value only inside the superconductor~\eqref{eq_J_normal_cyl}. This integral can readily be evaluated using the gauge potential at the shell, ${\boldsymbol{A}}(r = R) = B_{\rm n}/(2R) {\bf e}_\varphi$, the explicit expression for the strength of the magnetic field $B_{\rm n}$ given in Eq.~\eqref{eq_B_Omega}, and the normal current ${\boldsymbol{J}}_{\rm n}$ given in Eq.~\eqref{eq_J_normal_cyl}.}

\corr{The first term in the magnetic free energy~\eqref{eq_F_magn} is given by the integral over the whole volume of the cylinder and can be evaluated using the explicit expression for the generated magnetic field~\eqref{eq_B_Omega}. Adding the first and the third terms, we arrive at Eqs.~\eqref{eq_F_B} and \eqref{eq_I_magn} that we discuss below. The magnetostatic effect of the second term in the magnetic free energy~\eqref{eq_magn_B_E} will be addressed in Section~\ref{sec_free_dipole}.}

The overall minus sign in the free energy in Eq.~\eqref{eq_F_B} is expected: it is consistent with the standard form of the rotational contribution~\eqref{eq_F_I} of the mechanical rotation to the thermodynamic potential in the corotating reference frame at a fixed angular velocity $\boldsymbol{\Omega}$. The magnetic sector also yields a positive contribution to the inertial response, $I_{\rm magn}\equiv I_{B}>0$. This effect is understandable on physical grounds: when the surface charges of the normal component co-rotate with the shell, the induced current generates the magnetic field ${\boldsymbol{B}}_{\rm n}$ that stores energy. The field energy~\eqref{eq_F_B}, enters the free energy in the very same quadratic form in $\Omega$ as the purely mechanical rotational term does. Consequently, the electromagnetic and mechanical contributions add up, and the total moment of inertia of the system reads as follows:
\begin{align}
  I(\psi) = I_{\rm mech}(\psi) + I_{\rm magn}(\psi)\,,
  \label{eq_I_mech_I_magn}
\end{align}
where the mechanical moment of inertia is given in Eqs.~\eqref{eq_I_n}, \eqref{eq_I_n0} and \eqref{eq_I_s}, while the magnetic part of the moment of inertia is shown in Eq.~\eqref{eq_I_magn}. Both of these contributions depend on the condensate $\psi$, which specifies the decrease~\eqref{eq_I_s} of the moment of inertia due to the decoupling of electrons in the Cooper pairs and, respectively, the increase of the moment of inertia due to the energy stored in the generated magnetic field~\eqref{eq_B_Omega}.

\subsubsection{Effect of the moment of inertia on superconductivity}

Before proceeding to a more rigorous assessment of the consequences caused by rotation on superconductivity, let us first make a simple qualitative estimation. The key observation is that both mechanical and magnetic contributions to the total moment of inertia~\eqref{eq_I_mech_I_magn} depend on the density of Cooper pairs~\eqref{eq_n_s}. To simplify our analysis, we consider the condensate $\psi$ of a spatially uniform magnitude $|\psi(\boldsymbol{x})| = |\psi|$, which is valid for our geometry of a thin superconducting shell. 

Combining all contributions~\eqref{eq_I_n0}, \eqref{eq_I_s} and \eqref{eq_I_magn}, we get the total moment of inertia~\eqref{eq_I_mech_I_magn}:
\begin{align}
I(\psi) = \underbrace{I_{\rm n0} \frac{}{\!} \! - 2 m |\psi|^2 R^2 V_{\rm s}}_{{\rm mechanical\ rotation}} \, + \, \underbrace{\frac{8 \pi e^2}{c^2} d R \, |\psi|^4 R^2 V_{\rm s}}_{{\rm magnetic\ energy}}\,.
\label{eq_I_rot_psi}
\end{align}
The rotational part of the free energy at a fixed angular frequency~$\Omega$ is:
\begin{align}
	F_{\rm rot}(\psi) = - \frac{1}{2} I(\psi) \Omega^2\,.
    \label{eq_F_rot}
\end{align}
In thermal equilibrium, an isolated system achieves a minimum of its free energy as a function of internal parameters. Consequently, the rotating system~\eqref{eq_F_rot} tends to increase its moment of inertia at a fixed angular frequency $\Omega$.\footnote{An appropriate intuitive example is given by a glass of water spun at a fixed angular velocity about its symmetry axis. The water is pushed outward by the effective centrifugal force. It therefore redistributes toward the rim, leaving the surface lower near the axis and higher at the walls. Thus, the outward shift increases the moment of inertia of the system subjected to an externally fixed angular velocity.} 

In a rotating cylindrical superconducting shell, the single internal parameter of the system is the magnitude of the superconducting condensate $|\psi|$. The first, purely mechanical term in the total moment of inertia~\eqref{eq_I_rot_psi} tends to disfavor the formation of the condensate since the increasing condensate reduces the mechanical moment of inertia. However, the magnetic contribution given by the second term in~\eqref{eq_I_rot_psi} works in exactly the opposite way by favoring the increase of the superconducting condensate because the elevated condensate increases the energy stored in the magnetic field, which increases the moment of inertia of the rotating cylindrical shell. Thus, the fate of the superconductivity depends on a subtle interplay between these two factors that we discuss below. 

\subsubsection{Free energy of magnetostatic dipole interaction}
\label{sec_free_dipole}

The second term in the magnetic free energy~\eqref{eq_magn_B_E} can naturally be interpreted as the magnetostatic energy,
\begin{align}
    E_{\rm dipole} = &\, - \frac{1}{4 \pi} \int d^3 x \, {\boldsymbol{B}}_{\rm n} \cdot {\boldsymbol{H}_{\rm ext}}
    \equiv - {\boldsymbol{\mu}} \cdot {\boldsymbol{H}}_{\rm ext}
    \label{eq_E_dipole}\\
    = &\, - \frac{e}{c} \int_{V_{\rm s}} d^3 x \, \bigl(({\boldsymbol{H}}_{\rm ext} \times {\boldsymbol{x}} ) \cdot ({\boldsymbol{\Omega}} \times {\boldsymbol{x}})\bigr) |\psi|^2\,,
    \nonumber
\end{align}
corresponding to the interaction of the background magnetic field~${\boldsymbol{H}}_{\rm ext}$ with the magnetic dipole moment of the rotating cylinder:
\begin{align}
	{\boldsymbol{\mu}} \, & = \frac{1}{2c} \int_{V_{\rm s}} d^3 x \, {\boldsymbol{x}} \times {\boldsymbol{J}}_{\rm n}({\boldsymbol{x}})\,. 
\end{align}
The magnetic moment is produced by the circulating electric current of the normal component~\eqref{eq_J_normal}. The dipole interaction~\eqref{eq_E_dipole} of the rotating cylinder with the background magnetic field affects the superconducting phase transition because the magnetic dipole energy~\eqref{eq_E_dipole} is a quadratic function of the order parameter~$\psi$.

\subsection{Total free energy}

The total free energy~\eqref{eq_F_total_all} is the sum of the contributions coming from (i) the original GL energy of the condensate~\eqref{eq_F_GL}; (ii) the mechanical kinetic energy, Eqs.~\eqref{eq_F_I} and \eqref{eq_I_s}, of the normal part (ions and unpaired electrons) of the rotating system; (iii) the free energy of the generated magnetic field~\eqref{eq_magn_B_E} that includes Eqs.~\eqref{eq_F_B}, \eqref{eq_I_magn} and the magnetostatic dipole interaction~\eqref{eq_E_dipole}. Upon discarding terms that do not depend on the superconducting order parameter $\psi$, we get:
\begin{align}
    F = \int_{V_{\rm s}} d^3 x \biggl\{\biggl[& \, \frac{1}{4 m} 
    \Bigl(\hbar {\boldsymbol{\nabla}} \theta + \frac{2 e}{c} {\boldsymbol{A}} \Bigr)^2 + 
    \alpha + m \bigl(\boldsymbol{\Omega} \times {\boldsymbol{x}} \bigr)^2  \nonumber\\
    & \, - \frac{e}{c} \bigl(({\boldsymbol{H}}_{\rm ext} \times {\boldsymbol{x}} ) \cdot ({\boldsymbol{\Omega}} 
    \times {\boldsymbol{x}})\bigr)\biggr] |\psi|^2 \nonumber\\
    &\, + \frac{1}{2} \Bigl(\beta - \frac{8 \pi e^2}{c^2} d R^3 \Omega^2\Bigr) |\psi|^4 \biggr\}\,.
\label{eq_F_total_psi}
\end{align} 

To proceed further, it is convenient to express the free energy~\eqref{eq_F_total_psi} in a dimensionless form. To this end, we introduce the dimensionless GL parameter,
\begin{align}
	{\bar{\alpha}} = \frac{\alpha}{|\alpha_0|}\,, \qquad {\bar{\alpha}} \geqslant - 1\,,
    \label{eq_bar_alpha}
\end{align}
normalized at its zero-temperature value, $\alpha_0 {\equiv} \alpha(T=0)$, implying ${\bar{\alpha}}(T = 0) = -1$. Phenomenologically, the coupling $\alpha$ is usually taken to be a linear function of temperature~\cite{LL9}:
\begin{align}
    {\bar\alpha} = \frac{T}{T_{c0}} - 1\,.
    \label{eq_alpha_T}
\end{align}
The quadratic coupling $\alpha$ vanishes at the critical temperature, ${\bar{\alpha}}(T = T_{c0}) = 0$, where $T = T_{c0}$ corresponds to the transition in the absence of rotation and magnetic field.

The penetration depth $\lambda_0$ and the coherence length $\xi_0$ at $T=0$ are, respectively, as follows:
\begin{align}
    \lambda_0^2 = \frac{m c^2 \beta_0}{8 \pi e^2 |\alpha_0|}\,,
    \ \qquad
    \xi_0^2 = \frac{\hbar^2}{4 m |\alpha_0|}\,,
    \label{eq_lambda_xi_0} 
\end{align}
where the parameter $\beta$ is treated as a constant parameter, $\beta = \beta_0 \equiv \beta(T=0)$~\cite{LL9}. 

We express the control thermodynamic parameters, the angular frequency $\Omega$ and the background magnetic field $H_{\rm ext}$, via the characteristic angular frequency~$\Omega_0$ and the magnetic field strength~$H_0$, respectively:
\begin{align}
    \bar\Omega &\, = \frac{\Omega}{\Omega_0}\,, 
    \ \qquad\ 
    \Omega_0 = \frac{\hbar}{2 m \xi_0 R}\,,
    \label{eq_Omega_0} \\
	{\bar H} &\, = \frac{H_{\rm ext}}{H_0}\,, 
    \qquad
	H_0 = \frac{\phi_0}{2 \pi R \xi_0} \equiv \frac{\hbar c}{2 e R \xi_0}\,.
    \label{eq_H_0}
\end{align}
The characteristic magnetic field~\eqref{eq_H_0} corresponds to a half of the London magnetic field~\eqref{eq_B_L} generated by a solid superconductor that rotates with the characteristic frequency~\eqref{eq_Omega_0}: 
\begin{align}
    H_0 = \frac{1}{2} B_{L}(\Omega = \Omega_0)\,.	
\end{align}
The factor $1/2$ is introduced for further convenience.

We also normalize the condensate $\psi$ to its zero-temperature value $\psi_0$ in a non-rotating system~\cite{Kittel2018}:
\begin{align}
    |\bar\psi|^2 = \frac{|\psi|^2}{|\psi_0|^2}\,,
    \qquad
    |\psi_0|^2 = \frac{|\alpha_0|}{\beta_0} \equiv \frac{m c^2}{8 \pi e^2} \frac{1}{\lambda_0^2}\,.
    \label{eq_psi_normalized}
\end{align}
Notice, that by definition, 
\begin{align}
    0 \leqslant |\bar\psi| \leqslant 1\,.
    \label{eq_psi_bounds}
\end{align}
The value $|\bar\psi|=1$ corresponds to the state in which all electrons are condensed into Cooper pairs.

Finally, we introduce the scale for the free energy,
\begin{align}
    F_0 = |\alpha_0| |\psi_0|^2 V_{\rm s} \equiv \gamma L_z \biggl(\frac{\phi_0}{4 \pi \xi_0}\biggr)^2 
    \equiv \frac{\hbar^2 Z_{\rm I} n_{\rm I}}{\corr{8} m \xi_0^2} V_{\rm s}\,,
    \label{eq_F0}
\end{align}
and define a dimensionless geometrical factor~$\gamma$:
\begin{align}
    \gamma = \frac{R d}{\lambda_0^2}\,.
    \label{eq_gamma}
\end{align}

Before continuing our analysis further, let us consider the quadratic part $F^{(2)} \propto |\psi|^2$ of the free energy~\eqref{eq_F_total_psi}, $F = F^{(2)} + O(|\psi|^4)$. In the dimensionless units, the quadratic part reads:
\begin{align}
       \frac{F^{(2)}}{F_0} = \biggl[\biggl({\frac{\xi_0}{R}}\biggr)^2
       \bigl(k + {\bar \phi}_B \bigr)^2 + {\bar \alpha} + {\bar \Omega}^2 - {\bar H} {\bar \Omega} \biggr] |{\bar \psi}|^2\,,
       \label{eq_F_2}
\end{align}
where the magnetic flux $\phi_B = \pi R^2 B$ threading the cylinder enters via the superfluid velocity~\eqref{eq_v_s} as ${\bar \phi}_B = \phi_B/\phi_0$ normalized by the elementary magnetic flux~\eqref{eq_phi_0}.

The first term in~\eqref{eq_F_2} corresponds to the kinetic energy of the circulating Meissner current. It leads to the Little-Parks oscillations of critical temperature~\cite{Little1962} since, in the thermodynamic ground state, the winding number $k \in {\mathbb Z}$ adjusts itself in such a way that $|k + {\bar \phi}_B| < 1$. The second term comes from the GL potential in Eq.~\eqref{eq_F_GL}, the third term reflects the negative moment of inertia of the Cooper pairs~\eqref{eq_I_s} and the fourth term is the electrostatic dipole interaction~\eqref{eq_E_dipole}.

For our range of parameters, all terms in the free energy~\eqref{eq_F_2} are of the order of unity, except for the first (kinetic) term. The kinetic contribution is suppressed by the second power of the ratio $\xi_0/R$, which, in our geometry, is very small, $\xi_0/R \ll 1$. Therefore, the kinetic term of the superconducting condensate can be safely neglected in the free energy of our system.

In other words, we ignore the kinetic energy of the superconducting component in the free energy because the superconducting current~\eqref{eq_J_superconducting_cyl} is much weaker than the normal current~\eqref{eq_J_normal_cyl}. 

Intuitively, this property can also be understood as the result of the decoupling of the superfluid component from the rotating lattice in a thin ($\xi_0 \ll R$) cylinder: for a practical range of angular velocities $\Omega \sim \Omega_0$, the superfluid velocity~\eqref{eq_v_s} is much slower than the velocity of the mechanical rotation~\eqref{eq_v_n}:
\begin{align}
    \frac{|J_{\rm s}|}{|J_{\rm n}|} = \frac{|v_{\rm s}|}{\Omega R} \leqslant \frac{\xi_0}{R} 
    \frac{1}{{\bar{\Omega}}} \ll 1\,.
    \label{eq_Jn_dominance}
\end{align}

Dropping the inessential kinetic term and disregarding the contributions to the free energy~\eqref{eq_F_total_psi} that do not depend on the superconducting condensate, we arrive at the normalized free energy $f = F/F_0$ given by a purely potential term that includes effects of the rotation and the background magnetic field:
\begin{subequations}
\begin{align}
	f(\psi; {\bar\Omega}, {\bar H}) = &\, a(T,{\bar\Omega}, {\bar H}) |\bar\psi|^2 + \frac{1}{2} b({\bar\Omega}) |\bar\psi|^4 \,, 
    \label{eq_f}\\
    a(T,{\bar\Omega}, {\bar H}) = &\, {\bar \alpha}(T) + {\bar \Omega}^2 - {\bar\Omega} {\bar H} \,, 
    \label{eq_a}\\
    b({\bar\Omega}) = &\, 1 - \gamma {\bar \Omega}^2\,.
    \label{eq_b}
\end{align}
\label{eq_f_ab}
\end{subequations}

Before going into a detailed analysis of the phase diagram of the system, let us summarize the physical meaning of all contributions to the free energy~\eqref{eq_f}. The first term in the quadratic coefficient~\eqref{eq_a} and the first term in the quartic coefficient~\eqref{eq_b} are the standard (normalized) terms of the GL potential~\eqref{eq_F_GL}. The second term in the quadratic contribution~\eqref{eq_a} represents the positive contribution from the rotational energy of the normal component, Eqs.~\eqref{eq_F_I}-\eqref{eq_I_s}. The third term is the magnetostatic dipole interaction energy of the circular normal currents of the rotating cylinder with the background magnetic field~\eqref{eq_E_dipole}. Finally, the second term in the quartic coefficient~\eqref{eq_b} is the contribution from the magnetic-field energy given by Eqs.~\eqref{eq_F_B} and \eqref{eq_I_magn}.

\section{Phase diagram}
\label{sec_phase}

\subsection{General phase structure: three phases}

In the absence of condensate of Cooper pairs, $\psi = 0$, the system resides in a normal phase with a vanishing value of the normalized free energy~\eqref{eq_f}, $f(0) = 0$. A superconducting phase $\psi \neq 0$ is energetically favored over the normal phase provided $f(\psi) < 0$.

For a positive value of the quartic coupling, $b>0$, the thermodynamic ground state is determined by the sign of the quadratic coefficient $a$: a positive value, $a>0$ corresponds to a normal phase, while the negative value, $a<0$, promotes the formation of the superconducting state. The transition between these phases occurs when
\begin{align}
	a(T,{\bar\Omega}, {\bar H}) = 0\,, \qquad b({\bar \Omega}) > 0\,.
    \label{eq_a_eq_0}
\end{align}
This transition is of a second order.

A sufficiently strong rotation can also make the quartic coefficient~\eqref{eq_b} negative, $b<0$, enriching the phase structure of the system. Notice that despite the fact that the free energy~\eqref{eq_f} is not bounded from below for $b<0$, the ground state is stable against the runaway of the condensate $\psi$ because the density of the Cooper pairs is constrained~\eqref{eq_psi_bounds}. The maximal value of the condensate, $|\bar\psi| = 1$, is achieved when all available electrons are bound into Cooper pairs.   

At a positive $a>0$ and sufficiently small quartic coupling, $b < - a$, the free energy~\eqref{eq_f} develops a metastable superconducting minimum in a fully-paired state, $|\bar\psi| = 1$. This local minimum becomes a true ground state, $f(1) < f(0) \equiv 0$, when the quartic coefficient decreases below the threshold value of $b = - 2 a$:
\begin{align}
	2 a(T,{\bar \Omega}, {\bar H}) + b({\bar\Omega})  = 0\,, \qquad a({\bar \Omega}, {\bar H}) > 0\,,
    \label{eq_b_eq_2}
\end{align}
This transition is first order as the system evolves from the metastable state $|\bar\psi| = 0$ to the stable state $|\bar\psi| = 1$.

\corr{Notice that our conclusions can be affected by the terms in the GL energy that have higher powers in condensate (e.g., $|\psi|^6$, $|\psi|^8$, etc.). These terms were omitted in our analysis as in the immediate vicinity of a continuous phase transition, their contributions are expected to be subleading.  As these higher-order terms may potentially become relevant far from the transition regions, they can be incorporated systematically by extending the minimal GL functional.}

Before discussing the effects of the rotation and the background magnetic field on the superconducting temperature, it is instructive to plot the phase diagram in the space of the normalized couplings $a$ and $b$ of the GL potential~\eqref{eq_f} taking into account the natural constraint on the condensate of Cooper pairs~\eqref{eq_psi_bounds}

\begin{figure}[!htb]
    \centering
    \includegraphics[width=0.85\linewidth]{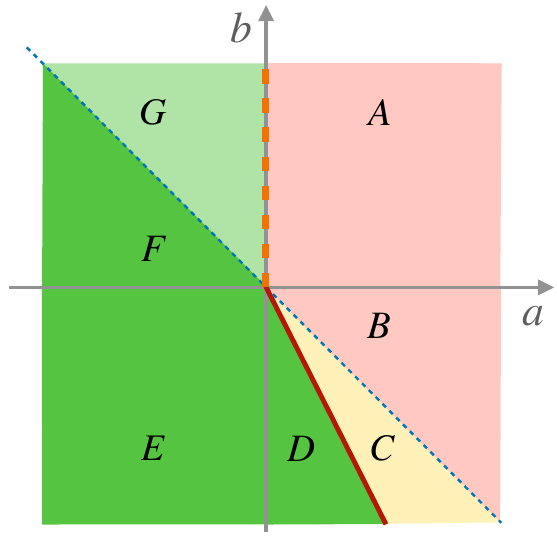}
    \caption{The phase diagram in the space of couplings $(a,b)$ of the free energy~\eqref{eq_f}. 
    The normal phase ($\psi = 0$) is shown in the light-red shading (the regions $A$ and $B$). In the yellow-shaded area $C$, the true ground state corresponds to the normal phase and the superconductivity appears as a metastable state. The light-green region $G$ points to the standard (in general, not fully paired) superconducting phase with $0 < |{\bar\psi}| < 1$. The saturated superconductivity with all electrons paired in condensed Cooper pairs $|{\bar\psi}| = 1$ emerges in the regions $D$, $E$ and $F$ shown by the darker green colors. The vertical dashed orange line $(a=0, b>0)$ marks the standard second-order phase transition~\eqref{eq_a_eq_0} between the normal phase $A$ with $(a>0,b>0)$ and the superconducting phase $G$ with $(a<0,b>0)$. The inclined solid red line $(2a + b =0, a>0)$ denotes the first-order phase transition~\eqref{eq_b_eq_2} between the metastable region $C$ and the superconducting phase $D$. The phases $B$ and $C$ as well as $F$ and $G$ are separated by the blue dashed line $a + b = 0$.}
    \label{fig_phase_ab}
\end{figure}

\begin{figure}[!htb]
    \centering
    \includegraphics[width=0.85\linewidth]{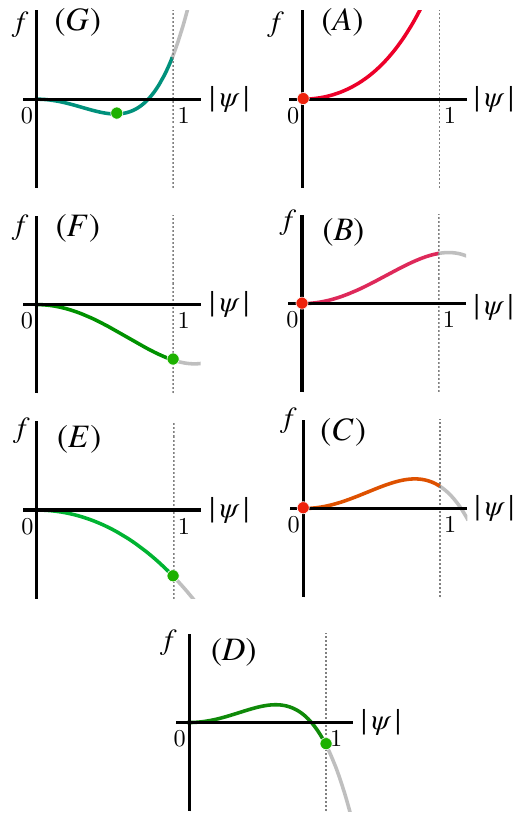}
    \caption{The schematic dependence of the free energy~\eqref{eq_f} on the amplitude of the superconducting order parameter $|\psi|$ in the regions $A$-$G$ of Fig.~\ref{fig_phase_ab}.}
    \label{fig_phase_potentials}
\end{figure}

The complete phase diagram in the space of couplings $(a,b)$ is shown in Fig.~\ref{fig_phase_ab}. The phase diagram can schematically be divided into seven regions, $A$, $B$, $\dots$, $G$, that are distinguished by a qualitative behavior of the thermodynamic potential~\eqref{eq_f} as shown in Fig.~\ref{fig_phase_potentials}. 

The system can reside in three phases. 
\begin{itemize}
    \item[(i)] {\bf Normal phase} appears as a single global minimum $\psi = 0$ of the free energy~\eqref{eq_f} in the regions $A$ ($b>0$) and $B$ ($b<0$) with
\begin{align}
    \left\{
    \begin{tabular}{r}
    $a>0$\,, \\
    \quad 
    $a + b > 0$\,,
	\end{tabular} 
    \right.
    \qquad 
    {\rm {[normal\ phase\ (\psi = 0)]}}\,.
\label{eq_normal_phase}
\end{align}
    The region $B$ has a common border with the region $C$, where the free energy has two minima: the normal phase as a global minimum supplemented by a metastable superconducting local minimum:
\begin{align}
	\left\{
    \begin{tabular}{c}
    $-2 a < b < - a$\,,\\
    $a > 0$ \,, 
    \end{tabular}
    \right.
    \qquad 
    \begin{tabular}{l}
    {\rm {[normal phase\ ($\psi = 0$)}} \\
    {\rm {with\ metastable\ supr.]}}
    \end{tabular}
    \label{eq_metastable}
\end{align}

    \item[(ii)] {\bf Fully paired superconducting phase} with all available electrons bound in condensed Cooper pairs (${\bar\psi} = 1$) in the regions $D$ ($a>0$, $b<0$), $E$ ($a<0$, $b<0$) and $F$ ($a<0$, $b>0$) that are constrained by the following relation:
\begin{align}
    b < {\rm{min}}\,(-2 a, - a)\,, \qquad 
    \begin{tabular}{l}
    {\rm {[saturated}}\ (${\bar \psi} = 1$) \\
    {\rm {superconductivity]}}
    \end{tabular}
    \label{eq_full_superconductivity}
\end{align}

    \item[(iii)] {\bf Usual superconducting state} in the region $G$:
\begin{align}
	\left\{
    \begin{tabular}{r}
	$a < 0$\,, \\
    $b + a > 0$\,, 
    \end{tabular}
    \right.
    \qquad 
        \begin{tabular}{r}
    {\rm {[ordinary ($0 < |\bar\psi| < 1$)}} \\
    {\rm {superconductivity]}}\,.
    \end{tabular}
\end{align}
    
\end{itemize}

The parameters of the normalized GL free energy~\eqref{eq_f_ab} depend nonlinearly on the angular frequency $\Omega$ and the background magnetic field $H_{\rm ext}$, thus making the phase diagram more involved. The precise phase structure depends not only on the characteristic frequency~\eqref{eq_Omega_0} and the characteristic magnetic field~\eqref{eq_H_0}, but also on the device-dependent geometric parameter~$\gamma$, Eq.~\eqref{eq_gamma} that enters the quartic coefficient~\eqref{eq_b}. Also, the quadratic coupling~\eqref{eq_a} depends on temperature~\eqref{eq_alpha_T}, which adds another parameter to the phase diagram. To estimate the effect of rotation on the phase structure of the system, we start our analysis with the simplest case of a relatively slow rotation in the absence of the background magnetic field.

\subsection{Enhancement of condensation by rotation}

\subsubsection{Qualitative picture at slow rotation}

Let us consider a specific case of a thin ($d \ll \lambda_0$) cylinder of a relatively large radius ($R \gg \lambda_0$) that rotates slowly ($\Omega \ll \Omega_0$) in the absence of a background magnetic field ($H_{\rm ext} = 0$). The dimensionless parameters~\eqref{eq_Omega_0}, \eqref{eq_H_0} and \eqref{eq_gamma} are as follows:
\begin{align}
    |{\bar \Omega}| \ll 1,
    \qquad
    {\bar H} = 0, 
    \qquad 
    \gamma \gg 1,
    \qquad
    \gamma {\bar \Omega}^2 \sim O(1).
    \label{eq_regime_H0}
\end{align}

At a vanishing magnetic field, ${\bar H} = 0$, the coupling constants of the free energy~\eqref{eq_f_ab} are simplified:
\begin{subequations}
\begin{align} 
    a(T,\Omega) = &\, \frac{T}{T_{c0}} - 1 + {\bar \Omega}^2
    \,, 
    \label{eq_a_H0}\\
    b(\Omega) = &\, 1 - \gamma {\bar \Omega}^2 \,,
    \label{eq_b_H0}
\end{align}
\label{eq_ab_H0}
\end{subequations}
\hskip -1mm where we used Eqs.~\eqref{eq_alpha_T} and \eqref{eq_Omega_0}. The effect of rotation is clearly visible in the couplings~\eqref{eq_ab_H0}. 

The physical meaning of the terms in Eq.~\eqref{eq_ab_H0} is transparent. A rotating system at a fixed angular velocity $\Omega$ tends to rearrange its internal degrees of freedom to increase its moment of inertia and to diminish the free energy~\eqref{eq_F_rot}. Thus, mechanically, rotation favors the normal electronic fraction---that adds to the moment of inertia---over the superconducting Cooper pairs that are decoupled from rotation. This effect breaks Cooper pairs and diminishes their density $|\psi|^2$ as reflected by the positive ${\bar \Omega}^2$ contribution to the quadratic coefficient~\eqref{eq_a_H0}. 

However, electromagnetically, the rotation works in an opposite way by favoring the emergence of the superconducting condensate that is decoupled from rotation. This effect increases the charge of the rotating normal component, which produces the circulating current proportional to the density of the Cooper pairs $|\psi|^2$. The current generates the magnetic field proportional to the current density squared, or to $|\psi|^4$. The magnetic field energy increases the moment of inertia of the system and diminishes the free energy~\eqref{eq_F_rot} thus contributing negatively to the quartic coupling~\eqref{eq_b_H0}. The magnetic field contribution ${\bar \Omega}^2$ comes with the geometric factor $\gamma$, Eq.~\eqref{eq_gamma}, since the magnetic field is produced in the whole volume inside the cylinder, which is much larger than the volume of the superconducting shell. 

Coming back to our discussion of the phase diagram, we disregard the last, $O({\bar\Omega}^2)$ term in the quadratic coupling~\eqref{eq_a_H0}, which is anyway negligibly small in the slow-rotation regime~\eqref{eq_regime_H0}. Then the phase diagram of the rotating system  is given precisely by Fig.~\ref{fig_phase_ab}, in which temperature $T$ controls the parameter $a$ via Eq.~\eqref{eq_a_H0}, while the angular velocity $\Omega$ determines the parameter $b$ as in Eq.~\eqref{eq_b_H0}.

Let us start with a nonrotating cylinder ($\Omega = 0$) in a normal state at a temperature $T > T_{c0}$. In terms of couplings~\eqref{eq_b_H0}, one has $a > 0$ and $b = 1$, so that the cylinder resides in the normal-state region $A$ of Fig.~\ref{fig_phase_ab}. 

As we increase rotation, at $\Omega = \Omega_0/\sqrt{\gamma}$ the quartic coupling vanishes,  $b = 0$, and a larger $\Omega$ it becomes negative. The system enters the other normal-state region $B$ that would be unstable if the condensate $|\psi|$ were unbounded from above~\eqref{eq_psi_bounds}. However, since the instability appears formally at $\bar\psi > 1$, unreachable by the system, the rotating system still resides in the non-superconducting phase~\eqref{eq_normal_phase} with $\psi = 0$.

As we increase the angular velocity higher, the coupling $b$ drops further, and the system enters the region $C$, where the maximally filled superconducting state $|\psi| = 1$ appears as a metastable local minimum of the free energy. However, the normal state $\psi = 0$ still remains the true thermodynamic ground state. 

Finally, as the rotation increases further, the system enters the region $D$, in which the maximal possible superconductivity, $\bar\psi = 1$, with all electrons paired in the condensed Cooper pairs~\eqref{eq_full_superconductivity}, becomes a true ground state. Let us stress again this surprising fact: the rotation forces a thin cylinder system in a normal state at temperature $T>T_{c0}$ to develop the maximally possible, strongest superconductivity. The latter phenomenon is usually realized only at the absolute zero temperature, $T = 0$. 

The critical line that separates the superconducting phase transition and the fully paired superconducting state is determined by Eq.~\eqref{eq_b_eq_2}, where the couplings $a$ and $b$ are given in Eq.~\eqref{eq_ab_H0}.

\begin{figure}[!htb]
    \centering
    \includegraphics[width=0.95\linewidth]{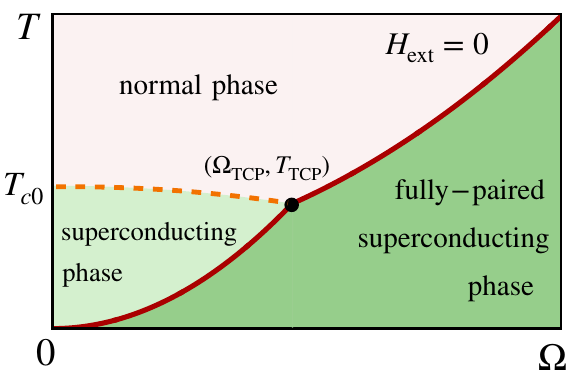}
    \caption{The schematic phase diagram of a thin cylindrical film rotating with the angular frequency $\Omega$ at the temperature $T$ in a vanishing background magnetic field $H_{\rm ext} = 0$. The geometric factor of the cylinder~\eqref{eq_gamma}, is greater than unity, $\gamma > 1$. A second-order phase transition line~\eqref{eq_Tc_II}, shown by the dashed orange line, separates the superconducting phase and the normal phase. At high rotation, the superconducting and normal phases turn, via a first-order phase transition~\eqref{eq_Tc_I}, shown by the solid red line, to the strong superconducting phase with all available electrons paired in the condensed Cooper pairs. The transition lines meet at the tricritical point~\eqref{eq_TCP}.}
    \label{fig_phase_T_Omega}
\end{figure}

\subsubsection{Phase diagram at vanishing magnetic field}

Let us now concentrate on a less constrained regime beyond a slow-rotation limit but still in the absence of a background magnetic field:
\begin{align}
H_{\rm ext} = 0\,, \qquad  \gamma \equiv \frac{R d}{\lambda_0^2} > 1\,.
\end{align}

Our previous analysis in the $(a,b)$ parameter space of the free energy~\eqref{eq_f_ab} shows that the regime of the usual superconductivity (the light green area in Fig.~\ref{fig_phase_ab}) \corr{appears in the range of couplings $b < 0$ and $a + b < 0$}. The fully paired superconductivity regime (the dark green area in Fig.~\ref{fig_phase_ab}) takes place at $b > 0$ with $a + b > 0$ and at $b < 0$ at $2a + b < 0$. Expressing the couplings $a$ and $b$ in terms of the temperature $T$ and the angular frequency $\Omega$ according to Eqs.~\eqref{eq_ab_H0}, we find the phase diagram shown in Fig.~\ref{fig_phase_T_Omega}.

At relatively high angular frequencies, the system appears in a fully paired phase that features the maximum superconductivity, $|\bar \psi| = 1$. As the angular frequency lowers, the superconducting state experiences the first-order phase transition either to the normal phase (at higher temperatures) or to the standard superconducting phase (at lower temperatures). The critical temperature $T_{c{\rm I}} = T_{c{\rm I}}(\Omega)$ of the first-order phase transition is the piecewise function of the angular frequency $\Omega$:
\begin{align}
	\frac{T_{c{\rm I}}(\Omega)}{T_{c0}} = 
    \left\{
    \begin{array}{rcl}
    (\gamma - 1) \bigl(\Omega/\Omega_0\bigr)^2 \,, & \qquad & |\Omega| \leqslant \Omega_{\rm TCP}\,, 
    \\[3mm]
    \!\! \frac{1}{2} + \bigl(\frac{\gamma}{2} - 1\bigr) \bigl(\Omega/\Omega_0\bigr)^2\,, 
    & \qquad & |\Omega| > \Omega_{\rm TCP}\,.
    \end{array}
    \right.
    \label{eq_Tc_I}
\end{align}

The transition line~\eqref{eq_Tc_I} exhibits a cusp at the tricritical point (TCP),
\begin{align}
	\biggl(\frac{\Omega_{\rm TCP}}{\Omega_0}, \frac{T_{\rm TCP}}{T_{c0}} \biggr) 
    = \biggl(\frac{1}{\sqrt{\gamma}}, \frac{\gamma - 1}{\gamma} \biggr)\,,
    \label{eq_TCP}
\end{align}
where it meets the second-order phase line:
\begin{align}
	\frac{T_{c{\rm II}}(\Omega)}{T_{c0}} = 1 - \frac{\Omega^2}{\Omega_0^2}\,, 
    \qquad
    |\Omega| \leqslant \Omega_{\rm TCP}\,,
    \label{eq_Tc_II}
\end{align}
which separates the lower-temperature pocket of the ordinary superconducting phase and the higher-temperature normal phase. 

Although in interesting and experimentally realizable scenarios the geometric factor~\eqref{eq_gamma} should be much larger than unity, $\gamma \gg 1$, let us briefly consider the phase diagram with $\gamma < 1$. 

One can quickly realize that at $\gamma < 1$, the fully paired phase disappears from the phase diagram because the first-order transition line~\eqref{eq_Tc_I} and the tricritical point~\eqref{eq_TCP} correspond to unphysical negative temperatures. The only transition that survives in this regime is the second-order transition line~\eqref{eq_Tc_II} between the normal and superconducting phases (the dashed orange line in Fig.~\ref{fig_phase_T_Omega}). As the angular velocity increases, the critical temperature diminishes parabolically because rotation elevates the density of the normal electronic fraction that favorably increases the moment of inertia of the system. The superconducting phase ceases to exist at $\Omega = \Omega_0$, where the critical temperature vanishes $T_{c{\rm I}}(\Omega = \Omega_0) = 0$. Hence, at a small geometric factor $\gamma < 1$ and a vanishing magnetic background field, $H_{\rm ext} = 0$, rotation inhibits superconductivity.

\subsubsection{Effect of a background magnetic field}

The presence of a magnetic field parallel to the axis of the rotating cylinder modifies the phase diagram. The quadratic coupling~\eqref{eq_a_H0} gets shifted by the magnetostatic dipole term that is linear in the magnetic strength $H_{\rm ext}$ and the angular frequency $\Omega$. This contribution leads to a shift of the critical temperatures in Eqs.~\eqref{eq_Tc_I} and \eqref{eq_Tc_II} by the same term ``$+(H_{\rm ext}/H_0) (\Omega/\Omega_0)$''.

In particular, the phenomenologically interesting second-order transition line between the ordinary superconducting phase and the higher-temperature normal phase~\eqref{eq_Tc_II} takes the following form:
\begin{align}
	\frac{T_{c{\rm II}}(\Omega)}{T_{c0}} = 1  + \frac{H_{\rm ext}}{H_0} \frac{\Omega}{\Omega_0}- \frac{\Omega^2}{\Omega_0^2} \qquad |\Omega| \leqslant \Omega_{\rm TCP}\,.
    \label{eq_Tc_II_H}
\end{align}
At the tricritical point, $\Omega_{\rm TCP} = \Omega_0/\sqrt{\gamma}$, the second-order transition merges with the first-order transition line.

The magnetic field co-aligned with the rotational velocity, $H_{\rm ext} \Omega > 0$, leads to an increase of the critical superconducting temperature, while the opposite orientation, $H_{\rm ext} \Omega < 0$, implies an inhibition of superconductivity. Similar selective rotational properties have been noticed in semiconductors of a comparable cylindrical/thin shell geometry~\cite{Chernodub:2021fpm}, although the consequences for semiconductors are much less dramatic than for superconductors. Below, we estimate the combined effect of magnetic field and rotation on the superconducting transition temperature under realistic conditions.

\section{Strength of the effect} 
\label{sec_strength}

Below, we estimate the effects of the rotation and the background magnetic field on the superconducting temperature. We will not discuss a concrete experimental setup to detect the superconductivity itself, noting that the superconductor transition can be found either directly, by measuring the drop in the resistivity of the film, or by illuminating a rotating cylinder made of a thin aluminum film with microwave photons and measuring their absorption coefficient, which serves as a reliable tool for the detection of the superconducting energy gap~\cite{Kittel2018, Biondi1956}. 

\subsection{Parameters of the system}

As an example, we consider a superconducting film made of pure aluminum (Al), which has an exceptionally long coherence length in the bulk, $\xi_{\rm Al} \simeq 1.6 \, \mu{\rm m}$~\cite{Kittel2018}. Notice that elemental tin (Sn) used in the original Little-Parks experiment has a much shorter coherence length, $\xi_{\rm Sn} \simeq 0.23\, \mu{\rm m}$~\cite{Kittel2018}. For sufficiently small thicknesses, $d \sim 50\,{\rm nm}$, the penetration length $\lambda$ of aluminum is larger than the width $d$~\cite{Lopez-Nunez:2023niv}, implying that the film satisfies the required conditions $d \lesssim \lambda_0$ and $ d \lesssim \xi_0$. 

The coherence length in thin films is usually shorter than in the bulk of the same materials~\cite{Tinkham1995}. For example, $\xi_0 \sim (10-100)\,{\rm nm}$ for $d \sim 50\,{\rm nm}$ aluminum films prepared via thermal evaporation~\cite{Steinberg2008, Liang2012}.  However, the superconducting coherence length in films depends also on their fabrication method. For example, in a film deposited with molecular beam epitaxy, the coherence length extends to $\xi_0 \simeq 8.86\,\mu{\rm m}$~\cite{Liang2012}. The local GL description is applicable for thin, sufficiently clean films~\cite{Tinkham1995, DeGennes2018}.

To estimate the strength of the effects, we choose a macroscopically large radius of the cylinder, $R = 1\,{\rm mm}$, which is substantially larger than the one ($\simeq 0.7\,\mu{\rm m}$) used in the Little-Parks experiment~\cite{Little1962}. Taking the coherence length $\xi_0 = 1\, \mu{\rm m}$, we get the characteristic rotation rate~\eqref{eq_Omega_0} 
\begin{align}
    \nu_0 = \frac{\Omega_0}{2\pi} \simeq 9.2\,{\rm kHz}\,,
    \label{eq_Omega0_system}
\end{align}
and the characteristic magnetic field~\eqref{eq_H_0}: 
\begin{align}
    H_0 = 3.3\times 10^{-3} \, {\rm G}\,.
    \label{eq_H0_system}
\end{align}

An aluminum film of the thickness $d \simeq 50\,{\rm nm}$ has the penetration depth $\lambda_0 \simeq 120\,{\rm nm}$~\cite{Lopez-Nunez:2023niv}, implying a very large value of the geometrical factor~\eqref{eq_gamma}:
\begin{align}
    \gamma \sim 3.5\times 10^3\,,
    \label{eq_gamma_system}
\end{align}
Equations~\eqref{eq_Omega0_system} and \eqref{eq_gamma_system} give us the frequency of the tricritical point~\eqref{eq_TCP}:
\begin{align}
    \nu_{\rm TCP} = \frac{\Omega_{\rm TCP}}{2\pi} \simeq 155\,{\rm Hz}\,.	
    \label{eq_Omega_TCP_system}
\end{align}

\subsection{Rotation in a magnetic field background}
\label{sec_rotation_magnetic}

Let us first discuss the combined effect of rotation and magnetic field on the second-order superconducting transition~\eqref{eq_Tc_II_H} between the standard superconducting phase and the normal phase.

Assuming that the cylinder rotates with the rotational rate $\nu = \Omega/(2 \pi) = 100\,{\rm Hz}$ at the background of the magnetic field $H_{\rm ext} = 10\, \rm{G}$ (both numbers do not seem outlandish), we get $\Omega/\Omega_0 \simeq 10^{-2}$ and $H_{\rm ext}/H_0 \simeq 3\times 10^3$. Equation~\eqref{eq_Tc_II_H} indicates that for these parameters, the purely rotational term $\Omega^2$ can be safely neglected, while the magnetostatic interaction raises the critical temperature by the large factor of 34. 

For a more general set of parameters, we define  
\begin{align}
    \Delta T_{c\rm{II}} = T_{c{\rm II}} - T_{c0} = (\Delta T_{c\rm{II}})_1 + (\Delta T_{c\rm{II}})_2\,,	
\end{align}
with $(\Delta T_c)_1/T_{c0} = (H_{\rm ext}/H_0) (\Omega/\Omega_0) $ and $(\Delta T_c)_2/T_{c0} = - (\Omega/\Omega_0)^2$, and get the following estimates:
\begin{align}
	\frac{(\Delta T_{c\rm{II}})_1}{T_{c0}}
    &\, {\simeq}\ 3.3 \times 
    {\biggl( \frac{H_{\rm ext}}{1\,{\rm G}} \biggr)}
    {\biggl( \frac{\nu}{1\,{\rm Hz}} \biggr)} 
    {\biggl( \frac{\xi_0}{1\,\mu{\rm m}} \biggr)}^2
    {\biggl( \frac{R}{1\,{\rm cm}} \biggr)}^2\!,
    \nonumber\\
    ~\label{eq_Delta_Tc}\\
	\frac{(\Delta T_{c\rm{II}})_2}{T_{c0}} 
    &\, {\simeq} - 1.2 \times 10^{-6} \times
    {\biggl( \frac{\nu}{1\,{\rm Hz}} \biggr)}^2 
    {\biggl( \frac{\xi_0}{1\,\mu{\rm m}} \biggr)}^2
    {\biggl( \frac{R}{1\,{\rm cm}} \biggr)}^2\!,
    \nonumber
\end{align}
where we used the rotational frequency $\nu = \Omega/(2 \pi)$. Notice that $\Omega$ and $H_{\rm ext}$ are, respectively, the $z$-components of the angular velocity and the background magnetic field that can also take negative values. For a parallel orientation of ${\boldsymbol{\Omega}}$ and ${\boldsymbol{H}}_{\rm ext}$, the critical temperature~\eqref{eq_Tc_II_H} increases, while it drops for an anti-parallel orientation of the rotation velocity and the magnetic field axis.

The superconducting critical temperature of a 50~nm--thick aluminum film is about $1.25$\,K~\cite{Chubov1969, Meservey1971, Ummarino2024}, implying that the rotation under these rather modest conditions should increase the critical temperature~\eqref{eq_Delta_Tc} to approximately $T_c \simeq 43$\,K. Thus, the magnetostatic dipole interaction~\eqref{eq_E_dipole} provides a substantial enhancement of superconductivity.

\subsection{Fully paired state via rotation}

Let us now consider the transition to the fully paired superconducting phase in which all electrons appear in the form of the condensed Cooper pairs, $|\bar\psi| = 1$. To reveal the effect of rotation, let us assume that the background magnetic field is absent, $H_{\rm ext} = 0$. Then, the enhancement of superconductivity is driven by the increase of the moment of inertia~\eqref{eq_I_magn} due to the energy stored in the magnetic field inside the cylinder~\eqref{eq_F_B}. The transition temperature is given by Eq.~\eqref{eq_Tc_I} and shown in Fig.~\ref{fig_phase_T_Omega} by the red solid line.

We take the frequency in a kHz band, $\Omega/(2\pi) \sim 1\,{\rm kHz}$, which is consistent with the range of parameters considered earlier~\eqref{eq_regime_H0}. In this regime: 
\begin{itemize}
    \item[(i)] The frequency $\Omega$ appears to be above the TCP value~\eqref{eq_Omega_TCP_system}, implying that the phase transition separates the fully paired superconductivity and the normal phase ({\it cf.} Fig.~\ref{fig_phase_T_Omega} at $\Omega > \Omega_{\rm TCP}$);

    \item[(ii)] the rotational correction---arising from the mechanical moment of inertia of the normal electron fraction  given by the last term in the quadratic GL coefficient~\eqref{eq_a_H0}---is small, $(\Omega/\Omega_0)^2 \simeq 10^{-2}$, and it will be neglected below; 
    
    \item[(iii)] the inertial contribution of the magnetic field energy---described by the $\Omega^2$ term in the quartic GL coefficient~\eqref{eq_b_H0}---is parametrically substantial, $\gamma (\Omega/\Omega_0)^2 \simeq 41$.
\end{itemize}

The critical temperature is given by the $\Omega > \Omega_{\rm TCP}$ branch of Eq.~\eqref{eq_Tc_I}. Since $\gamma \gg 1$, we neglect the additive $O(1)$ terms in Eq.~\eqref{eq_Tc_I} and get the following result for the shift of the critical temperature:
\begin{align}
    \Delta T_{c\rm{I}} = T_{c{\rm I}} - T_{c0}\,,	
\end{align}
where
\begin{align}
    \frac{\Delta T_{c\rm{I}}}{T_{c0}} \simeq 295 \times
    {\biggl( \frac{\nu}{1\,{\rm kHz}} \biggr)}^2 
    {\biggl( \frac{R}{1\,{\rm cm}} \biggr)}^3
    {\biggl( \frac{\xi_0}{\lambda_0} \biggr)}^2\,,
    \label{eq_Delta_T_II}
\end{align}
and we set $d = 50\,{\rm nm}$ in the calculation. For the thin cylindrical film of the radius $R = 1\,{\rm mm}$, with $\xi_0 = 1\, \mu{\rm m}$ and $\lambda_0 \simeq 120\,{\rm nm}$, the enhancement of the critical temperature at the rotational rate $\nu = 1\,{\rm kHz}$ is about 20, pointing to an increased superconducting temperature $T_c \simeq 25\, \mathrm{K}$.

The enhancement~\eqref{eq_Delta_T_II} becomes more pronounced as the ratio $\xi_0/\lambda_0$ increases. It is therefore expected to be stronger in type-I superconductors, for which $\xi_0 / \lambda_0 > \sqrt{2}$, as compared to type-II materials, that are characterized by smaller $\xi_0 / \lambda_0$. Bulk aluminum is a deep type-I superconductor with $\xi_0 / \lambda_0 \simeq 100$, making it a particularly favorable candidate for observing rotation-induced strengthening of superconductivity.

However, in thin films, the ratio $\xi_0 / \lambda_0$ can be much smaller as, in general, thin superconducting films exhibit reduced coherence lengths $\xi_0$ compared to the same materials in bulk. This effect originates primarily due to the surface scattering and reduced electron mean free paths since decreasing the film thickness $d$ eventually reduces the grain size in the material~\cite{Meservey1971}. 
The ratio may depend on the fabrication method that affects the cleanliness of the material and its granularity~\cite{Maloney1972, Steinberg2008, Liang2012, Lopez-Nunez:2023niv}. 

Still, the enhancement effect scales quadratically with the rotation rate and cubically with the radius of the cylinder~\eqref{eq_Delta_T_II}. Thus, raising one of these parameters by one order of magnitude amplifies the corresponding shift of the critical temperature by two or three orders, respectively. The strong $\Omega^2$ and $R^3$ scalings offer sufficient parameter leverage to offset a smaller $\xi_0 / \lambda_0$ ratio and maintain a large enhancement effect of the critical temperature even in less favorable materials.

\corr{In the parameter regime where rotation enhances superconductivity, the normal state, $\psi = 0$, is unstable. Any small superconducting fluctuation, whether of quantum or thermal origin, may seed the instability and drive the system into the superconducting phase. From an experimental perspective, however, it may be more appropriate to prepare the sample initially in the superconducting phase, set it into rotation, and then gradually increase the temperature to determine whether rotation raises the critical temperature.}

\subsection{Strengths of the electromagnetic fields}

Let us discuss the magnitude of the electromagnetic fields that are produced by the rotating cylinder.

\subsubsection{Magnetic field}

The magnetic field generated inside the cylinder~\eqref{eq_B_Omega} by rotation is
\begin{align}
B_\Omega = &\, \frac{8 \pi e}{c} |\psi|^2 R d \, \Omega
 \equiv \gamma \frac{m \phi_0 \Omega}{\pi \hbar} |{\bar \psi}|^2\,,
\label{eq_B_Omega_system_1}
\end{align}
where $\phi_0$ is the flux quantum~\eqref{eq_phi_0}. The maximal strength of the field~\eqref{eq_B_Omega_system_1} is achieved in a maximally paired state with $|{\bar \psi}| = 1$, which provides an upper bound on the magnetic field generated by rotation: 
\begin{align}
B_\Omega \simeq  \, 1.2 \times
    {\biggl( \frac{\nu}{1\,{\rm kHz}} \biggr)} \times 1 \, {\rm G}\,,
\label{eq_B_Omega_system_2}
\end{align}

Thus, the rotation at the rate $1\,{\rm kHz}$ produces the magnetic field with the strength~\eqref{eq_B_Omega_system_2} of the order of $B_{\Omega} \simeq 1 \,{\rm G}$, which is substantially weaker than the critical value~\cite{Kittel2018} $B_{c} \simeq 105 \, {\rm G}$ for the bulk aluminum at $T=0$. One should also notice that the critical value $B_{c\|}$ of the magnetic field tangential to the surface to the film is much higher than the bulk critical field $B_{c}$~\cite{Toxen1964}. For example, for an aluminum film of the thickness $d \sim 100\,{\rm nm}$, the critical magnetic field is $B_{c\|} \sim 10^4 \, {\rm G}$~\cite{Meservey1971}. Therefore, the magnetic field generated by rotation remains well below the relevant pair-breaking scale. In other words, the rotation-driven onset of superconductivity is not hindered by the self-generated magnetic field.

For reference, the London magnetic field~\eqref{eq_B_L} at the characteristic rotation rate $\nu_0 = \Omega_0/(2\pi) \simeq 9.2\,{\rm kHz}$ has a much smaller value: $B_L \equiv 2 H_0 \simeq 6.6 \times 10^{-3}\, {\rm G}$. The magnetic field of such strength would have been produced in a rotating solid (thick) superconducting cylinder. The weakness of the London field comes from the fact that the solid cylinder also generates a strong screening Meissner current of the superconducting component. In the thin shell, such screening currents are negligible and the magnetic field is not screened, hence $B_\Omega \gg B_L$.

\corr{In realistic experimental settings, one can also expect the presence of a magnetic-field component, which is normal to the superconducting cylindrical shell. This residual radial component of the background magnetic field may lead to a flux-trapping effect in the superconducting film. Although the presence of vortices may be compatible with the mechanism proposed in our article, a quantitative analysis of their influence lies beyond the scope of the present work, as it requires a more detailed analysis. For an experimental realization of the proposed mechanism, magnetic-field compensation and shielding may be essential.}

\subsubsection{Energy balance: magnetic field vs. condensation energy}

One could ask the following puzzling question. The magnetic field generated by rotation~\eqref{eq_B_Omega_system_2} has a relatively weak strength $B_{\Omega} \simeq 1 \,{\rm G}$, which is much smaller than the critical magnetic field~$B_{c} \simeq 105 \, {\rm G}$ of the superconducting transition. How can the physics associated with the generation of such a weak magnetic field, $B_{\Omega} \ll B_c$, be a meaningful reason for the substantial increase of superconducting transition temperature in the film? 

The answer is that the rotation-produced magnetic field~\eqref{eq_B_Omega_system_2} emerges in the whole interior of the cylinder, which has a much larger volume compared to the volume of the thin superconducting film. Therefore, the energy of the system associated with the weak magnetic field generated in the large volume of the cylinder competes with the strong condensation energy in the thin superconducting film of a much smaller volume. 

Let us put these considerations into numbers using the physical parameters of our example described above. The cylinder rotating at the rate $\nu = 1\,{\rm kHz}$ generates the magnetic field~\eqref{eq_B_Omega_system_2} with the strength $B_{\rm \Omega} \simeq 1.3\,{\rm G}$, which has the following energy density per volume:
\begin{align}
    U_{\rm B} = \frac{1}{8\pi} B_{\Omega}^2 \simeq 6.7\times 10^{-9}\,{\rm J}/{\rm cm}^3	\,.
    \label{eq_U_estimation}
\end{align}
(for convenience, we give the result in the SI units). The magnetic field is generated in the whole interior volume of the cylinder. Therefore, the energy density of the magnetic field per unit length of the cylinder is: 
\begin{align}
	{\mathcal{U}}_{\rm B} \equiv \frac{U_{\rm B} V_{\rm cyl}}{L_z} = \pi R^2 U_{\rm B} \simeq 2.1\times 10^{-10}\,{\rm J}/{\rm cm}\,,
    \label{eq_E_cylinder}
\end{align}
where $V_{\rm cyl} = \pi R^2 L_z$ is the volume of the interior of the cylinder (we recall that, in our example, $R = 1\, {\rm mm}$).

At the same time, the free energy density associated with the condensation energy of Cooper pairs in the bulk aluminum is~\cite{Kittel2018}:
\begin{align}
	{(\Delta F)}_{\rm supr} = \frac{1}{8 \pi} B_c^2 \simeq &\, 430\,{\rm erg}/{\rm cm}^3 \nonumber\\ 
    \equiv &\, 4.3 \times 10^{-5}\,{\rm J}/{\rm cm}^3\,,
    \label{eq_Delta_F_estimation}
\end{align}
with $1\,{\rm erg} = 10^{-7}\,{\rm J}$. Expectedly, ${(\Delta F)}_{\rm supr} \gg U_{\rm B}$. 

The condensation of Cooper pairs occurs only in the thin film. Therefore, the condensation energy per unit length of the cylinder is:
\begin{align}
	{(\Delta {\mathcal F})}_{\rm supr} \equiv \frac{{(\Delta F)}_{\rm supr} V_{\rm s}}{L_z}
    \simeq 1.4 \times 10^{-10}\,{\rm J}/{\rm cm}\,.
    \label{eq_E_film}
\end{align}
where $V_{\rm s} = 2\pi R d L_z$ is the volume of the cylindrical film (the thickness of our cylinder is $d = 50\,{\rm nm}$).

Remarkably, the energy densities per unit length of the cylinder, associated with the magnetic field~\eqref{eq_E_cylinder} and the condensation~\eqref{eq_E_film}, are of the same order! The difference in the scales of the energy densities per unit volume, magnetic~\eqref{eq_U_estimation} vs. superconducting~\eqref{eq_Delta_F_estimation}, is leveraged by the large geometrical factor $V_{\rm cyl}/V_{\rm s} = R/(2d) = 10^4$. 

Therefore, the decrease in the free energy~\eqref{eq_F_B} due to the generation of the magnetic field~\eqref{eq_B_Omega}, has the same magnitude as the gain in the free energy due to (otherwise unfavorable) condensation of Cooper pairs. In other words, rotation enhances condensation in the thin cylindrical film via the generation of the magnetic field in the whole interior of the cylinder.

\subsubsection{Electric field}

Motion in the background magnetic field ${\boldsymbol{B}}$ produces the electric field ${\boldsymbol{E}} = - {\boldsymbol{v}} \times {\boldsymbol{B}}$ in the comoving frame. For rotational motion, ${\boldsymbol{v}} = {\boldsymbol{\Omega}} \times {\boldsymbol{x}}$, and the strength of the field $E = - B R \Omega$. For the set of parameters mentioned in our example, the induced electric field is tiny, $|E| \simeq 6 \times 10^{-6} {\rm V}/ {\rm cm}$.

The centrifugal force acting on an electron of the normal, non-condensed fraction,
\begin{align}
	{\boldsymbol{F}}_{\rm cf} = - m {\boldsymbol{\Omega}} \times ({\boldsymbol{\Omega} \times {\boldsymbol{R}}}) \equiv m \Omega^2 R \, {\bf e}_\rho\,,
\end{align}
leads to a voltage drop 
\begin{align}
    \Delta V = \frac{F_{\rm cf} d}{e} = \frac{\hbar^2 {\bar \Omega}^2}{4 m e \xi_0^2} \frac{d}{R}\,, 
\end{align}
along the radial direction $\rho$ across the film. For our set of parameters, this effect is also negligible, $\Delta V \sim 10^{-16} {\rm V}$.

\corr{As we mentioned earlier, one can also expect the presence of a radial magnetic-field component in a realistic experimental setting. This magnetic component, which is normal to the superconducting cylindrical shell, should generate a small in-plane longitudinal electric field that will, in turn, produce a weak longitudinal supercurrent. The typical order of magnitude of these effects is much smaller than a typical Cooper-pair breaking scale of about $E \sim 10^6\, {\rm V}/{\rm cm}$ in the films~\cite{Ruf2024, Zaccone2025}. Therefore, the effects of the generated electric fields can be safely neglected.}

\section{Conclusions} 

We argue that rotation of a thin superconducting cylindrical shell about its symmetry axis can significantly enhance the critical temperature of the superconducting transition. Our arguments are based on the observation that in a rotating shell, electrons in condensed Cooper pairs decouple from rotation: the circulating Meissner current of condensed electrons is much weaker compared to the electric current produced by the rotating normal component of ions and normal electron fraction. 

The basic idea is as follows. In a normal state, a negative electric charge of normal electrons fully compensates a positive electric charge of the ionic lattice. Thus, because of its electric neutrality, a rotating shell in the normal state does not generate a magnetic field at all. However, the condensation of Cooper pairs eliminates a part of the normal electrons from the rotating shell so that the charge compensation in the rotating component is no longer complete. Therefore, the rotating cylinder produces a circulating electric current of charged ions and a remaining normal electron fraction. This current, in turn, generates a magnetic dipole moment directed along the axis of the rotating cylinder. The stronger the condensate, the larger the magnetic field generated in the cylinder and, consequently, the larger its magnetic dipole moment. The generated magnetic field gives rise to two distinct effects that strengthen superconductivity.

The first enhancement effect occurs in the presence of a background magnetic field. The magnetostatic interaction between the background field and the magnetic dipole moment of the rotating cylinder favors an increase in the dipole moment of the rotating cylinder, which, in turn, promotes the condensation of Cooper pairs.

The second enhancement effect takes place in the absence of the external field. A thermodynamic argument implies that, at fixed angular velocity, the system tends to maximize its moment of inertia. In our system, this effect arises from an increase in the strength of the generated magnetic field: the rotational energy is stored not only in the mechanical motion but also in the magnetic field produced due to rotation, thus supporting the condensation of Cooper pairs.

We estimated these effects in a cylinder made of a thin aluminum film and pointed out their experimental feasibility. Both effects increase the critical temperature by an order of magnitude.

\begin{acknowledgments}
MC is partially supported by the EU’s NextGenerationEU instrument through the National Recovery and Resilience Plan of Romania - Pillar III-C9-I8, managed by the Ministry of Research, Innovation and Digitization, within the project FORQ, contract no.~760079/23.05.2023 code CF 103/15.11.2022. FW is supported by the U.S. Department of Energy under grant Contract Number DE-SC0012567 and by the Swedish Research Council under Contract No. 335-2014-7424. The authors are grateful to G.~E.~Volovik for useful correspondence.
\end{acknowledgments}

\appendix 

\section{Comments on earlier works}
\label{sec_comments}

The critical temperature of a rotating thin superconducting cylinder in a background magnetic field can also be inferred, in part, from the analyses reported in Refs.~\cite{Verkin1972, Capellmann2002}. However, the underlying Ginzburg–Landau formulations employed in those works are not identical to the approach adopted in our article (or to each other). The most important difference arises because these analyses fail to include the magnetic field generated by the normal state charge imbalance. We therefore devote this Appendix to a critical examination of the origin of the apparent discrepancies and outline precisely which assumptions and model ingredients are responsible for the quantitative and qualitative mismatches between the two approaches.

\subsection{Comment on Ref.~\cite{Verkin1972}}

Starting in the historical order, we first discuss Ref.~\cite{Verkin1972}. There are two essential differences between our works.

\subsubsection{Magnetic (dipole) interaction}

The magnetic dipole coupling between the normal component of the rotating thin cylinder and the background magnetic field---represented by the fourth term of the total free energy~\eqref{eq_F_total_psi}---is absent in Ref.~\cite{Verkin1972}. The reason for the omission is that the GL equations in Ref.~\cite{Verkin1972} were stated a priori rather than derived variationally from an explicit free-energy functional.

This difference is crucial since it is the missed magnetic dipole coupling that drives the enhancement of the critical temperature by rotation in the presence of a background magnetic field~\eqref{eq_Tc_II_H}. 

Consequently, the effect of the condensation on the generation of the magnetic field in the interior of the cylinder has also not been taken into account in Ref.~\cite{Verkin1972}. This effect is crucial since the free energy of the magnetic field~\eqref{eq_F_B}, which enters as the last term of the total free energy~\eqref{eq_F_total_psi}, drives the enhancement of the critical temperature by rotation in the absence of a background magnetic field~\eqref{eq_Tc_I}.

\subsubsection{Subleading effect: Kinetic energy}

The kinetic term of Cooper pairs, in Eq.~(3.7) of Ref.~\cite{Verkin1972}, depends only on the difference, $({\boldsymbol{v}}_{\rm s} - {\boldsymbol{v}}_{\rm n})^2$, of the velocities of the superconducting component~\eqref{eq_v_s}, ${\boldsymbol{v}}_{\rm s}$, and the normal component~\eqref{eq_v_n}, ${\boldsymbol{v}}_{\rm n}$. The appearance of this term might seem to be well-justified from the point of view of Galilean invariance. Indeed, for inertial motion, the internal energy of a physical body does not depend on the velocity of the body with respect to any given inertial frame. This requirement implies the invariance of the internal energy under the change in all velocities by a constant vector $\boldsymbol{u}$. For example, in a two-fluid picture of superconductivity, the change of the reference frame results in the shifts ${\boldsymbol{v}}_{\rm s} \to {\boldsymbol{v}}_{\rm s} - {\boldsymbol{u}}$ and ${\boldsymbol{v}}_{\rm n} \to {\boldsymbol{v}}_{\rm n} - {\boldsymbol{u}}$ that leave the difference in velocities, ${\boldsymbol{v}}_{\rm s} - {\boldsymbol{v}}_{\rm n}$, invariant.

On the contrary, the free energy used in our paper includes the sum of the individual kinetic energies, proportional to ${\boldsymbol{v}}_{\rm s}^2$ and $ {\boldsymbol{v}}_{\rm n}^2$, as represented by the first and the third terms in Eq.~\eqref{eq_F_total_psi}. These terms violate Galilean invariance.

The breaking of Galilean invariance in rotating systems appears due to the non-inertial nature of rotation. Clearly, the internal energy, similarly to the centrifugal force, can depend on the absolute value of the angular velocity $\boldsymbol{\Omega}$ with $\boldsymbol{\Omega} = 0$ defining a physically distinguished (non-rotating) reference frame. 

The origin of the difference in the kinetic terms of Ref.~\cite{Verkin1972} and our work can also be traced back to our earlier observation that the superfluid component in a thin superconducting cylinder detaches mechanically from the rotating ionic lattice. Our point of view is also shared by the analysis of Ref.~\cite{Capellmann2002}, which argues that the angular momentum of the superconducting condensate, ${\boldsymbol{l}}_{\rm s} = {\boldsymbol{x}} \times {\boldsymbol{p}}_{\rm s} \equiv 2 m {\boldsymbol{x}} \times {\boldsymbol{v}}_{\rm s} $, should not couple mechanically to rotation. In other words, the Lagrange multiplier $- {\boldsymbol{\Omega}} \cdot {\boldsymbol{l}}_{\rm s}$ should not be included in the expression of the free energy in the corotating reference frame, in line with the Legendre transformation in Eq.~\eqref{eq_F_I_general}. Physically, the decoupling means that the rotation of the lattice has no direct mechanical influence on the kinetic energy of Cooper pairs.\footnote{The rotation still affects the strength of the phonon coupling of Cooper pairs that leads to a mass renormalization, $m \to m^*$ mentioned in footnote~\ref{foot_phonons} on page~\pageref{foot_phonons}. This is a subleading effect.}

We emphasize that the subtlety associated with defining the kinetic energy of Cooper pairs is specific to thin superconducting shells, in which the superfluid component is not mechanically coupled to the motion of the ionic lattice. In the bulk of a thick superconductor, the superfluid component is rigidly dragged by the ionic lattice and, therefore, our arguments are no longer valid. 

This remark is primarily of formal interest, since the precise form of the kinetic coupling is inessential for our analysis: the kinetic energy of the Cooper pairs gives a negligible contribution to the free energy compared to other terms.

\subsection{Comment on Ref.~\cite{Capellmann2002}}

A closely related analysis has also been performed in Ref.~\cite{Capellmann2002}, where the GL free energy of a rotating superconductor in a magnetic-field background has been derived. Despite the superconducting phase transition in an external magnetic field has not been addressed in Ref.~\cite{Capellmann2002}, we can still make a comparison of our approaches.

Similarly to the case of Ref.~\cite{Verkin1972}, the difference between the free energies in our papers appears in the omission of the magnetic dipole energy in Ref.~\cite{Capellmann2002}. The origin of this omission can be traced to the Maxwell law (in the notations of Ref.~\cite{Capellmann2002}) ${\boldsymbol{\nabla}} \times {\boldsymbol{H}} = (4\pi/c) {\boldsymbol{J}}$, which describes the magnetic field ${\boldsymbol{H}}$ generated by the circulating electric current ${\boldsymbol{J}}$ of the rotating normal component. In Ref.~\cite{Capellmann2002}, the normal current has been associated only with the ionic current density, ${\boldsymbol{J}} \to {\boldsymbol{J}}_{\rm I} = e Z_{\rm I} n_{\rm I} \boldsymbol{v}_{\rm I}$, which does not depend on the condensate. Therefore, the magnetic field ${\boldsymbol{H}}$ has been treated in Ref.~\cite{Capellmann2002} as an external parameter.

However, the normal component of the system also contains a contribution from the normal-state electrons, which rotate together with the ionic lattice. The negatively charged normal fraction of electrons partially screens the positive electric charge of the ions, implying that the source current should contain both contributions: ${\boldsymbol{J}} \to {\boldsymbol{J}}_{\rm n} = e Z_{\rm I} n_{\rm I} \boldsymbol{v}_{\rm I} - e n_{{\rm n}} \boldsymbol{v}_{\rm n}$. 

As we mentioned above, the local electric neutrality condition relates the normal electric current ${\boldsymbol{J}}_{\rm n}$ to the density $|\psi|^2$ of the Cooper pairs~\eqref{eq_J_normal}, implying
${\boldsymbol{J}} \to {\boldsymbol{J}}_{\rm n} = 2 e ({\boldsymbol{\Omega}} \times {\boldsymbol{x}}) |\psi({\boldsymbol{x}})|^2$. Consequently, the right-hand side of the Maxwell equation in Ref.~\cite{Capellmann2002} should be corrected as follows: ${\boldsymbol{\nabla}} \times {\boldsymbol{H}} = (8 e\pi/c) ({\boldsymbol{\Omega}} \times {\boldsymbol{x}}) |\psi({\boldsymbol{x}})|^2$. The magnetic field ${\boldsymbol{H}}$ generated by the normal component of the rotating lattice is thus tightly related to the density of Cooper pairs, and, therefore, the field ${\boldsymbol{H}}$ cannot be treated as a {\it fixed} background magnetic field. This observation appears to be crucial for our analysis.

By an explicit calculation, one can show---following the line of derivation of Ref.~\cite{Capellmann2002}---that this missed contribution corresponds precisely to the magnetic dipole interaction represented by the fourth term in Eq.~\eqref{eq_F_total_psi}. Moreover, the magnetic field generated by the neutral current in the interior of the cylinder---given in the last term in the total free energy~\eqref{eq_F_total_psi}---has consequently been omitted. Our derivation of these missed terms has been presented in Section~\ref{sec_free_magnetic_energy}.


\begin{thebibliography}{27}%
\makeatletter
\providecommand \@ifxundefined [1]{%
 \@ifx{#1\undefined}
}%
\providecommand \@ifnum [1]{%
 \ifnum #1\expandafter \@firstoftwo
 \else \expandafter \@secondoftwo
 \fi
}%
\providecommand \@ifx [1]{%
 \ifx #1\expandafter \@firstoftwo
 \else \expandafter \@secondoftwo
 \fi
}%
\providecommand \natexlab [1]{#1}%
\providecommand \enquote  [1]{``#1''}%
\providecommand \bibnamefont  [1]{#1}%
\providecommand \bibfnamefont [1]{#1}%
\providecommand \citenamefont [1]{#1}%
\providecommand \href@noop [0]{\@secondoftwo}%
\providecommand \href [0]{\begingroup \@sanitize@url \@href}%
\providecommand \@href[1]{\@@startlink{#1}\@@href}%
\providecommand \@@href[1]{\endgroup#1\@@endlink}%
\providecommand \@sanitize@url [0]{\catcode `\\12\catcode `\$12\catcode `\&12\catcode `\#12\catcode `\^12\catcode `\_12\catcode `\%12\relax}%
\providecommand \@@startlink[1]{}%
\providecommand \@@endlink[0]{}%
\providecommand \url  [0]{\begingroup\@sanitize@url \@url }%
\providecommand \@url [1]{\endgroup\@href {#1}{\urlprefix }}%
\providecommand \urlprefix  [0]{URL }%
\providecommand \Eprint [0]{\href }%
\providecommand \doibase [0]{http://dx.doi.org/}%
\providecommand \selectlanguage [0]{\@gobble}%
\providecommand \bibinfo  [0]{\@secondoftwo}%
\providecommand \bibfield  [0]{\@secondoftwo}%
\providecommand \translation [1]{[#1]}%
\providecommand \BibitemOpen [0]{}%
\providecommand \bibitemStop [0]{}%
\providecommand \bibitemNoStop [0]{.\EOS\space}%
\providecommand \EOS [0]{\spacefactor3000\relax}%
\providecommand \BibitemShut  [1]{\csname bibitem#1\endcsname}%
\let\auto@bib@innerbib\@empty
\bibitem [{\citenamefont {London}(1961)}]{London1961}%
  \BibitemOpen
  \bibfield  {author} {\bibinfo {author} {\bibfnamefont {F.}~\bibnamefont {London}},\ }\href@noop {} {\emph {\bibinfo {title} {Superfluids, Volume I: Macroscopic Theory of Superconductivity}}}\ (\bibinfo  {publisher} {Dover Publications},\ \bibinfo {address} {New York},\ \bibinfo {year} {1961})\ \bibinfo {note} {originally published by Wiley in 1950}\BibitemShut {NoStop}%
\bibitem [{\citenamefont {Becker}\ \emph {et~al.}(1933)\citenamefont {Becker}, \citenamefont {Heller},\ and\ \citenamefont {Sauter}}]{Becker1933}%
  \BibitemOpen
  \bibfield  {author} {\bibinfo {author} {\bibfnamefont {R.}~\bibnamefont {Becker}}, \bibinfo {author} {\bibfnamefont {G.}~\bibnamefont {Heller}}, \ and\ \bibinfo {author} {\bibfnamefont {F.}~\bibnamefont {Sauter}},\ }\bibfield  {title} {\enquote {\bibinfo {title} {{\"Uber die Stromverteilung in einer supraleitenden Kugel} ({On} the current distribution in a superconducting sphere)},}\ }\href {\doibase 10.1007/bf01330324} {\bibfield  {journal} {\bibinfo  {journal} {Zeitschrift f\"ur Physik}\ }\textbf {\bibinfo {volume} {85}},\ \bibinfo {pages} {772–787} (\bibinfo {year} {1933})}\BibitemShut {NoStop}%
\bibitem [{\citenamefont {Capellmann}(2002)}]{Capellmann2002}%
  \BibitemOpen
  \bibfield  {author} {\bibinfo {author} {\bibfnamefont {H.}~\bibnamefont {Capellmann}},\ }\bibfield  {title} {\enquote {\bibinfo {title} {Rotating superconductors: {Ginzburg-Landau} equations},}\ }\href {\doibase 10.1140/e10051-002-0004-z} {\bibfield  {journal} {\bibinfo  {journal} {The European Physical Journal B - Condensed Matter}\ }\textbf {\bibinfo {volume} {25}},\ \bibinfo {pages} {25–30} (\bibinfo {year} {2002})}\BibitemShut {NoStop}%
\bibitem [{\citenamefont {Little}\ and\ \citenamefont {Parks}(1962)}]{Little1962}%
  \BibitemOpen
  \bibfield  {author} {\bibinfo {author} {\bibfnamefont {W.~A.}\ \bibnamefont {Little}}\ and\ \bibinfo {author} {\bibfnamefont {R.~D.}\ \bibnamefont {Parks}},\ }\bibfield  {title} {\enquote {\bibinfo {title} {Observation of quantum periodicity in the transition temperature of a superconducting cylinder},}\ }\href@noop {} {\bibfield  {journal} {\bibinfo  {journal} {Physical Review Letters}\ }\textbf {\bibinfo {volume} {9}},\ \bibinfo {pages} {9} (\bibinfo {year} {1962})}\BibitemShut {NoStop}%
\bibitem [{\citenamefont {De~Gennes}(2018)}]{DeGennes2018}%
  \BibitemOpen
  \bibfield  {author} {\bibinfo {author} {\bibfnamefont {P.~G.}\ \bibnamefont {De~Gennes}},\ }\href {\doibase 10.1201/9780429497032} {\emph {\bibinfo {title} {Superconductivity of Metals and Alloys}}}\ (\bibinfo  {publisher} {CRC Press},\ \bibinfo {year} {2018})\BibitemShut {NoStop}%
\bibitem [{\citenamefont {Verkin}\ and\ \citenamefont {Kulik}(1972)}]{Verkin1972}%
  \BibitemOpen
  \bibfield  {author} {\bibinfo {author} {\bibfnamefont {B.~I.}\ \bibnamefont {Verkin}}\ and\ \bibinfo {author} {\bibfnamefont {I.~O.}\ \bibnamefont {Kulik}},\ }\bibfield  {title} {\enquote {\bibinfo {title} {Magnetic fields of rotating superconductors},}\ }\href {http://jetp.ras.ru/cgi-bin/e/index/e/34/5/p1103?a=list} {\bibfield  {journal} {\bibinfo  {journal} {Sov. Phys. JETP}\ }\textbf {\bibinfo {volume} {34}},\ \bibinfo {pages} {1103} (\bibinfo {year} {1972})}\BibitemShut {NoStop}%
\bibitem [{\citenamefont {Berger}(2004)}]{Berger2004}%
  \BibitemOpen
  \bibfield  {author} {\bibinfo {author} {\bibfnamefont {J.}~\bibnamefont {Berger}},\ }\bibfield  {title} {\enquote {\bibinfo {title} {Nonlinearity of the field induced by a rotating superconducting shell},}\ }\href {http://dx.doi.org/10.1103/PhysRevB.70.212502} {\bibfield  {journal} {\bibinfo  {journal} {Physical Review B}\ }\textbf {\bibinfo {volume} {70}} (\bibinfo {year} {2004})}\BibitemShut {NoStop}%
\bibitem [{\citenamefont {Lipavský}\ \emph {et~al.}(2013)\citenamefont {Lipavský}, \citenamefont {Bok},\ and\ \citenamefont {Koláček}}]{Lipavsk2013}%
  \BibitemOpen
  \bibfield  {author} {\bibinfo {author} {\bibfnamefont {P.}~\bibnamefont {Lipavský}}, \bibinfo {author} {\bibfnamefont {J.}~\bibnamefont {Bok}}, \ and\ \bibinfo {author} {\bibfnamefont {J.}~\bibnamefont {Koláček}},\ }\bibfield  {title} {\enquote {\bibinfo {title} {Time-dependent {Ginzburg-Landau} equations for rotating and accelerating superconductors},}\ }\href {\doibase 10.1016/j.physc.2013.06.007} {\bibfield  {journal} {\bibinfo  {journal} {Physica C: Superconductivity}\ }\textbf {\bibinfo {volume} {492}},\ \bibinfo {pages} {144–151} (\bibinfo {year} {2013})}\BibitemShut {NoStop}%
\bibitem [{\citenamefont {Babaev}\ and\ \citenamefont {Svistunov}(2014)}]{Babaev2014}%
  \BibitemOpen
  \bibfield  {author} {\bibinfo {author} {\bibfnamefont {E.}~\bibnamefont {Babaev}}\ and\ \bibinfo {author} {\bibfnamefont {B.}~\bibnamefont {Svistunov}},\ }\bibfield  {title} {\enquote {\bibinfo {title} {Rotational response of superconductors: {Magnetorotational} isomorphism and rotation-induced vortex lattice},}\ }\href {http://dx.doi.org/10.1103/PhysRevB.89.104501} {\bibfield  {journal} {\bibinfo  {journal} {Phys. Rev. B}\ }\textbf {\bibinfo {volume} {89}} (\bibinfo {year} {2014})}\BibitemShut {NoStop}%
\bibitem [{\citenamefont {Hirsch}(2019{\natexlab{a}})}]{Hirsch2019a}%
  \BibitemOpen
  \bibfield  {author} {\bibinfo {author} {\bibfnamefont {J.E.}\ \bibnamefont {Hirsch}},\ }\bibfield  {title} {\enquote {\bibinfo {title} {Moment of inertia of superconductors},}\ }\href {\doibase 10.1016/j.physleta.2018.09.031} {\bibfield  {journal} {\bibinfo  {journal} {Physics Letters A}\ }\textbf {\bibinfo {volume} {383}},\ \bibinfo {pages} {83–90} (\bibinfo {year} {2019}{\natexlab{a}})}\BibitemShut {NoStop}%
\bibitem [{\citenamefont {Hirsch}(2019{\natexlab{b}})}]{Hirsch2019b}%
  \BibitemOpen
  \bibfield  {author} {\bibinfo {author} {\bibfnamefont {J.~E.}\ \bibnamefont {Hirsch}},\ }\bibfield  {title} {\enquote {\bibinfo {title} {Defying inertia: How rotating superconductors generate magnetic fields},}\ }\href {http://dx.doi.org/10.1002/andp.201900212} {\bibfield  {journal} {\bibinfo  {journal} {Annalen der Physik}\ }\textbf {\bibinfo {volume} {531}} (\bibinfo {year} {2019}{\natexlab{b}})}\BibitemShut {NoStop}%
\bibitem [{\citenamefont {Braguta}\ \emph {et~al.}(2024)\citenamefont {Braguta}, \citenamefont {Chernodub}, \citenamefont {Roenko},\ and\ \citenamefont {Sychev}}]{Braguta:2023yjn}%
  \BibitemOpen
  \bibfield  {author} {\bibinfo {author} {\bibfnamefont {V.}~\bibnamefont {Braguta}}, \bibinfo {author} {\bibfnamefont {M.}~\bibnamefont {Chernodub}}, \bibinfo {author} {\bibfnamefont {A.}~\bibnamefont {Roenko}}, \ and\ \bibinfo {author} {\bibfnamefont {D.}~\bibnamefont {Sychev}},\ }\bibfield  {title} {\enquote {\bibinfo {title} {{Negative moment of inertia and rotational instability of gluon plasma}},}\ }\href {\doibase 10.1016/j.physletb.2024.138604} {\bibfield  {journal} {\bibinfo  {journal} {Phys. Lett. B}\ }\textbf {\bibinfo {volume} {852}},\ \bibinfo {pages} {138604} (\bibinfo {year} {2024})}\BibitemShut {NoStop}%
\bibitem [{\citenamefont {Kittel}\ and\ \citenamefont {McEuen}(2018)}]{Kittel2018}%
  \BibitemOpen
  \bibfield  {author} {\bibinfo {author} {\bibfnamefont {Ch.}\ \bibnamefont {Kittel}}\ and\ \bibinfo {author} {\bibfnamefont {P.}~\bibnamefont {McEuen}},\ }\href@noop {} {\emph {\bibinfo {title} {Introduction to solid state physics}}}\ (\bibinfo  {publisher} {John Wiley \& Sons},\ \bibinfo {year} {2018})\BibitemShut {NoStop}%
\bibitem [{\citenamefont {Landau}\ and\ \citenamefont {Lifshitz}(1996)}]{LL9}%
  \BibitemOpen
  \bibfield  {author} {\bibinfo {author} {\bibfnamefont {L.~D.}\ \bibnamefont {Landau}}\ and\ \bibinfo {author} {\bibfnamefont {E.~M.}\ \bibnamefont {Lifshitz}},\ }\href@noop {} {\emph {\bibinfo {title} {Statistical Physics, Part 2}}},\ \bibinfo {edition} {3rd}\ ed.\ (\bibinfo  {publisher} {Butterworth-Heinemann},\ \bibinfo {address} {Oxford, England},\ \bibinfo {year} {1996})\BibitemShut {NoStop}%
\bibitem [{\citenamefont {Chernodub}(2021)}]{Chernodub:2021fpm}%
  \BibitemOpen
  \bibfield  {author} {\bibinfo {author} {\bibfnamefont {M.~N.}\ \bibnamefont {Chernodub}},\ }\bibfield  {title} {\enquote {\bibinfo {title} {{Rotational diode: Clockwise/counterclockwise asymmetry in conducting and mechanical properties of rotating (semi)conductors}},}\ }\href {\doibase 10.3390/sym13091569} {\bibfield  {journal} {\bibinfo  {journal} {Symmetry}\ }\textbf {\bibinfo {volume} {13}},\ \bibinfo {pages} {1569} (\bibinfo {year} {2021})},\ \Eprint {http://arxiv.org/abs/2104.05032} {arXiv:2104.05032} \BibitemShut {NoStop}%
\bibitem [{\citenamefont {Biondi}\ \emph {et~al.}(1956)\citenamefont {Biondi}, \citenamefont {Garfunkel},\ and\ \citenamefont {McCoubrey}}]{Biondi1956}%
  \BibitemOpen
  \bibfield  {author} {\bibinfo {author} {\bibfnamefont {M.~A.}\ \bibnamefont {Biondi}}, \bibinfo {author} {\bibfnamefont {M.~P.}\ \bibnamefont {Garfunkel}}, \ and\ \bibinfo {author} {\bibfnamefont {A.~O.}\ \bibnamefont {McCoubrey}},\ }\bibfield  {title} {\enquote {\bibinfo {title} {Millimeter wave absorption in superconducting aluminum},}\ }\href {\doibase 10.1103/physrev.101.1427.2} {\bibfield  {journal} {\bibinfo  {journal} {Physical Review}\ }\textbf {\bibinfo {volume} {101}},\ \bibinfo {pages} {1427–1429} (\bibinfo {year} {1956})}\BibitemShut {NoStop}%
\bibitem [{\citenamefont {L{\'o}pez-N{\'u}{\~n}ez}\ \emph {et~al.}(2025)\citenamefont {L{\'o}pez-N{\'u}{\~n}ez}, \citenamefont {Torras-Coloma}, \citenamefont {Montserrat}, \citenamefont {Bertoldo}, \citenamefont {Cozzolino}, \citenamefont {Rius}, \citenamefont {Mart{\'\i}nez},\ and\ \citenamefont {Forn-D{\'\i}az}}]{Lopez-Nunez:2023niv}%
  \BibitemOpen
  \bibfield  {author} {\bibinfo {author} {\bibfnamefont {David}\ \bibnamefont {L{\'o}pez-N{\'u}{\~n}ez}}, \bibinfo {author} {\bibfnamefont {Alba}\ \bibnamefont {Torras-Coloma}}, \bibinfo {author} {\bibfnamefont {Queralt~Portell}\ \bibnamefont {Montserrat}}, \bibinfo {author} {\bibfnamefont {Elia}\ \bibnamefont {Bertoldo}}, \bibinfo {author} {\bibfnamefont {Luca}\ \bibnamefont {Cozzolino}}, \bibinfo {author} {\bibfnamefont {Gemma}\ \bibnamefont {Rius}}, \bibinfo {author} {\bibfnamefont {M.}~\bibnamefont {Mart{\'\i}nez}}, \ and\ \bibinfo {author} {\bibfnamefont {P.}~\bibnamefont {Forn-D{\'\i}az}},\ }\bibfield  {title} {\enquote {\bibinfo {title} {{Magnetic penetration depth of Aluminum thin films}},}\ }\href {\doibase 10.1088/1361-6668/adf360} {\bibfield  {journal} {\bibinfo  {journal} {Supercond. Sci. Technol.}\ }\textbf {\bibinfo {volume} {38}},\ \bibinfo {pages} {095004} (\bibinfo {year} {2025})},\ \Eprint {http://arxiv.org/abs/2311.14119} {arXiv:2311.14119} \BibitemShut {NoStop}%
\bibitem [{\citenamefont {Tinkham}(1995)}]{Tinkham1995}%
  \BibitemOpen
  \bibfield  {author} {\bibinfo {author} {\bibfnamefont {Michael}\ \bibnamefont {Tinkham}},\ }\href@noop {} {\emph {\bibinfo {title} {Introduction to Superconductivity}}},\ \bibinfo {edition} {2nd}\ ed.\ (\bibinfo  {publisher} {McGraw-Hill},\ \bibinfo {address} {New York},\ \bibinfo {year} {1995})\BibitemShut {NoStop}%
\bibitem [{\citenamefont {Steinberg}\ \emph {et~al.}(2008)\citenamefont {Steinberg}, \citenamefont {Scheffler},\ and\ \citenamefont {Dressel}}]{Steinberg2008}%
  \BibitemOpen
  \bibfield  {author} {\bibinfo {author} {\bibfnamefont {K.}~\bibnamefont {Steinberg}}, \bibinfo {author} {\bibfnamefont {M.}~\bibnamefont {Scheffler}}, \ and\ \bibinfo {author} {\bibfnamefont {M.}~\bibnamefont {Dressel}},\ }\bibfield  {title} {\enquote {\bibinfo {title} {Quasiparticle response of superconducting aluminum to electromagnetic radiation},}\ }\href {http://dx.doi.org/10.1103/PhysRevB.77.214517} {\bibfield  {journal} {\bibinfo  {journal} {Physical Review B}\ }\textbf {\bibinfo {volume} {77}} (\bibinfo {year} {2008})}\BibitemShut {NoStop}%
\bibitem [{\citenamefont {Liang}\ \emph {et~al.}(2012)\citenamefont {Liang}, \citenamefont {Yeh}, \citenamefont {Lin}, \citenamefont {Wu}, \citenamefont {Lin}, \citenamefont {Chen} \emph {et~al.}}]{Liang2012}%
  \BibitemOpen
  \bibfield  {author} {\bibinfo {author} {\bibfnamefont {C.-T.}\ \bibnamefont {Liang}}, \bibinfo {author} {\bibfnamefont {M.-R.}\ \bibnamefont {Yeh}}, \bibinfo {author} {\bibfnamefont {S.W.}\ \bibnamefont {Lin}}, \bibinfo {author} {\bibfnamefont {J.Y.}\ \bibnamefont {Wu}}, \bibinfo {author} {\bibfnamefont {T.L.}\ \bibnamefont {Lin}}, \bibinfo {author} {\bibfnamefont {Kuang~Yao}\ \bibnamefont {Chen}},  \emph {et~al.},\ }\bibfield  {title} {\enquote {\bibinfo {title} {Superconductivity in an aluminum film grown by molecular beam epitaxy},}\ }\href@noop {} {\bibfield  {journal} {\bibinfo  {journal} {Chinese Journal of Physics}\ }\textbf {\bibinfo {volume} {50}},\ \bibinfo {pages} {638--642} (\bibinfo {year} {2012})}\BibitemShut {NoStop}%
\bibitem [{\citenamefont {Chubov}\ \emph {et~al.}(1969)\citenamefont {Chubov}, \citenamefont {Eremenko},\ and\ \citenamefont {Pilipenko}}]{Chubov1969}%
  \BibitemOpen
  \bibfield  {author} {\bibinfo {author} {\bibfnamefont {P.~N.}\ \bibnamefont {Chubov}}, \bibinfo {author} {\bibfnamefont {V.~V.}\ \bibnamefont {Eremenko}}, \ and\ \bibinfo {author} {\bibfnamefont {Yu.~A.}\ \bibnamefont {Pilipenko}},\ }\bibfield  {title} {\enquote {\bibinfo {title} {Dependence of the critical temperature and energy gap on the thickness of superconducting aluminum films},}\ }\href@noop {} {\bibfield  {journal} {\bibinfo  {journal} {Sov Phys JETP}\ }\textbf {\bibinfo {volume} {28}},\ \bibinfo {pages} {389--395} (\bibinfo {year} {1969})}\BibitemShut {NoStop}%
\bibitem [{\citenamefont {Meservey}\ and\ \citenamefont {Tedrow}(1971)}]{Meservey1971}%
  \BibitemOpen
  \bibfield  {author} {\bibinfo {author} {\bibfnamefont {R.}~\bibnamefont {Meservey}}\ and\ \bibinfo {author} {\bibfnamefont {P.~M.}\ \bibnamefont {Tedrow}},\ }\bibfield  {title} {\enquote {\bibinfo {title} {Properties of very thin aluminum films},}\ }\href {\doibase 10.1063/1.1659648} {\bibfield  {journal} {\bibinfo  {journal} {J. Appl. Phys.}\ }\textbf {\bibinfo {volume} {42}},\ \bibinfo {pages} {51–53} (\bibinfo {year} {1971})}\BibitemShut {NoStop}%
\bibitem [{\citenamefont {Ummarino}\ and\ \citenamefont {Zaccone}(2024)}]{Ummarino2024}%
  \BibitemOpen
  \bibfield  {author} {\bibinfo {author} {\bibfnamefont {G.~A.}\ \bibnamefont {Ummarino}}\ and\ \bibinfo {author} {\bibfnamefont {A.}~\bibnamefont {Zaccone}},\ }\bibfield  {title} {\enquote {\bibinfo {title} {Quantitative eliashberg theory of the superconductivity of thin films},}\ }\href {\doibase 10.1088/1361-648x/ad92ed} {\bibfield  {journal} {\bibinfo  {journal} {Journal of Physics: Condensed Matter}\ }\textbf {\bibinfo {volume} {37}},\ \bibinfo {pages} {065703} (\bibinfo {year} {2024})}\BibitemShut {NoStop}%
\bibitem [{\citenamefont {Maloney}\ \emph {et~al.}(1972)\citenamefont {Maloney}, \citenamefont {de~la Cruz},\ and\ \citenamefont {Cardona}}]{Maloney1972}%
  \BibitemOpen
  \bibfield  {author} {\bibinfo {author} {\bibfnamefont {M.~D.}\ \bibnamefont {Maloney}}, \bibinfo {author} {\bibfnamefont {F.}~\bibnamefont {de~la Cruz}}, \ and\ \bibinfo {author} {\bibfnamefont {M.}~\bibnamefont {Cardona}},\ }\bibfield  {title} {\enquote {\bibinfo {title} {Superconducting parameters and size effects of aluminum films and foils},}\ }\href {\doibase 10.1103/physrevb.5.3558} {\bibfield  {journal} {\bibinfo  {journal} {Physical Review B}\ }\textbf {\bibinfo {volume} {5}},\ \bibinfo {pages} {3558–3572} (\bibinfo {year} {1972})}\BibitemShut {NoStop}%
\bibitem [{\citenamefont {Toxen}(1964)}]{Toxen1964}%
  \BibitemOpen
  \bibfield  {author} {\bibinfo {author} {\bibfnamefont {A.~M.}\ \bibnamefont {Toxen}},\ }\bibfield  {title} {\enquote {\bibinfo {title} {Temperature dependence of the critical fields of thin superconducting films},}\ }\href {\doibase 10.1103/revmodphys.36.308} {\bibfield  {journal} {\bibinfo  {journal} {Rev. Mod. Phys.}\ }\textbf {\bibinfo {volume} {36}},\ \bibinfo {pages} {308} (\bibinfo {year} {1964})}\BibitemShut {NoStop}%
\bibitem [{\citenamefont {Ruf}\ \emph {et~al.}(2024)\citenamefont {Ruf}, \citenamefont {Puglia}, \citenamefont {Elalaily}, \citenamefont {De~Simoni}, \citenamefont {Joint}, \citenamefont {Berke} \emph {et~al.}}]{Ruf2024}%
  \BibitemOpen
  \bibfield  {author} {\bibinfo {author} {\bibfnamefont {L.}~\bibnamefont {Ruf}}, \bibinfo {author} {\bibfnamefont {C.}~\bibnamefont {Puglia}}, \bibinfo {author} {\bibfnamefont {T.}~\bibnamefont {Elalaily}}, \bibinfo {author} {\bibfnamefont {G.}~\bibnamefont {De~Simoni}}, \bibinfo {author} {\bibfnamefont {F.}~\bibnamefont {Joint}}, \bibinfo {author} {\bibfnamefont {M.}~\bibnamefont {Berke}},  \emph {et~al.},\ }\bibfield  {title} {\enquote {\bibinfo {title} {Gate control of superconducting current: Mechanisms, parameters, and technological potential},}\ }\href {\doibase 10.1063/5.0222371} {\bibfield  {journal} {\bibinfo  {journal} {Applied Physics Reviews}\ }\textbf {\bibinfo {volume} {11}},\ \bibinfo {pages} {041314} (\bibinfo {year} {2024})}\BibitemShut {NoStop}%
\bibitem [{\citenamefont {Zaccone}\ \emph {et~al.}(2025)\citenamefont {Zaccone}, \citenamefont {Ummarino}, \citenamefont {Braggio},\ and\ \citenamefont {Giazotto}}]{Zaccone2025}%
  \BibitemOpen
  \bibfield  {author} {\bibinfo {author} {\bibfnamefont {Alessio}\ \bibnamefont {Zaccone}}, \bibinfo {author} {\bibfnamefont {Giovanni~Alberto}\ \bibnamefont {Ummarino}}, \bibinfo {author} {\bibfnamefont {Alessandro}\ \bibnamefont {Braggio}}, \ and\ \bibinfo {author} {\bibfnamefont {Francesco}\ \bibnamefont {Giazotto}},\ }\bibfield  {title} {\enquote {\bibinfo {title} {Microscopic mechanism of electric-field-induced superconductivity suppression in metallic thin films},}\ }\href {\doibase 10.1103/PhysRevB.111.174528} {\bibfield  {journal} {\bibinfo  {journal} {Phys. Rev. B}\ }\textbf {\bibinfo {volume} {111}},\ \bibinfo {pages} {174528} (\bibinfo {year} {2025})}\BibitemShut {NoStop}%
\end{thebibliography}
\end{document}